\documentclass[%
 reprint,
 amsmath,amssymb,mathrsfs,
 aps,
 longbibliography,
]{revtex4-1}

\usepackage{graphicx}
\usepackage{dcolumn}
\usepackage{bm}

\begin{document}

\preprint{APS/123-QED}

\title{Plasmonic lenses for tunable ultrafast electron emitters at the nanoscale}

\author{Daniel B. Durham$^{1,2}$, Fabrizio Riminucci$^{3,4}$, Filippo Ciabattini$^{5}$, Andrea Mostacci$^{3}$, Andrew M. Minor$^{1,2}$, Stefano Cabrini$^{5}$, and Daniele Filippetto$^{4}$}
\email{dfilippetto@lbl.gov; }
\affiliation{$^{1}$ Department of Materials Science and Engineering,
University of California, Berkeley, 
Berkeley, California 94720, USA \\
$^{2}$ National Center for Electron Microscopy, Molecular Foundry, Lawrence Berkeley National Laboratory, One Cyclotron Road, Berkeley, California, 94720, USA\\
$^{3}$ University of Rome "La Sapienza", Piazzale Aldo Moro 5, 00185 Rome, Italy\\
$^{4}$ Accelerator Technology and Applied Physics Division, Lawrence Berkeley National Laboratory, One Cyclotron Road, Berkeley, California, 94720, USA \\
$^{5}$ Molecular Foundry, Lawrence Berkeley National Laboratory, One Cyclotron Road, Berkeley, California, 94720, USA\\}

\date{\today}

\begin{abstract}

Simultaneous spatio-temporal confinement of energetic electron pulses to femtosecond and nanometer scales is a topic of great interest in the scientific community,  given the potential impact of such development on a wide spectrum of scientific and industrial applications. For example, in ultrafast electron scattering, nanoscale probes would enable accurate maps of structural dynamics in materials with nanoscale heterogeneity, thereby understanding the role of boundaries and defects on macroscopic properties. On the other hand, advances in this field are mostly limited by the electron source brightness and size.
We present the design, fabrication, and optical characterization of bullseye plasmonic lenses for next-generation ultrafast electron sources.  Using electromagnetic simulations, we examine how the interplay between light-plasmon coupling, plasmon propagation, dispersion, and resonance governs the properties of the photoemitted electron pulse. We also illustrate how the pulse duration and strength can be tuned by geometric design, and predict sub-10 fs pulses with nanoscale diameter can be achieved. We then fabricated lenses in gold films and characterized their plasmonic properties with cathodoluminescence spectromicroscopy, demonstrating suitable plasmonic behavior for ultrafast, nanoscale photoemission. 

\end{abstract}

\maketitle

\section{\label{sec:level1}Introduction}
Ultrafast electron sources have been extensively used as tools for scientific discovery over the past two decades. Such sources can now reliably produce electron pulses with femtosecond duration, which can be used as probes for dynamic microscopy and scattering measurements. Ultrafast electron-based measurements have provided insight into mechanisms of structural phase transitions in condensed matter~\cite{siwick2003AlMelting,siwick2014VO2,dwaynemiller2010TaS2} as well as chemical reactions and photochemistry in gases and molecular solids~\cite{zewail2006review,dwaynemiller2013MITOrganicSalt}. 
Even so, ultrafast electron experimentation is still limited by the source brightness, defined as the number density of electrons in transverse phase space [i.e. per unit solid angle and unit area, also called four-dimensional (4D) emittance]~\cite{brightness}, and setting a limit to beam relative coherence and focusability~\cite{UEDperspective_2012}. Spatio-temporal mapping with both nanoscale spatial and femtosecond temporal resolution is rarely utilized~\cite{ropers2018UCBED,ropers2018ULTEM,zewail2012UEELSPlasmonic}, and typically requires a combination of high contrast signals, several hour acquisition times, and limited sampling. Electron-based mapping of dynamics with increased throughput and detail may provide key insights into how microstructure and defects locally influence phase transformation, carrier generation and recombination, phonon and plasmon propagation, and much more. 

In effort to increase brightness, researchers have developed radio-frequency-based electron guns (rf guns) capable of sustaining large electric field amplitudes at extraction and during acceleration. By providing an order of magnitude higher accelerating field (100 MV/m) than direct-current electron guns, the extracted peak current can be increased by orders of magnitude, from the $\mu$A-range typical of electron microscopes to several A, while generating relativistic electron pulses with sub-10-fs duration~\cite{maxson_direct_2017,zhao_terahertz_2018}. Yet, tip emitters typically used in electron microscopes are not easily used in rf guns since under such high fields, they have limited lifetime and tend to emit dark electrons via field emission~\cite{DarkCurrentApex}, which are not synchronous with the laser pulse and add background to the experiment. Instead, flat cathodes are often used, which have much larger emission areas in the range of tens of $\mu$m or larger, ultimately limited by the incident laser spot size. Considering conservation of the beam emittance, larger source size in turn produces proportionally larger spot size for a given angular spread, thereby limiting the minimum useful spot size for applications.

Recently, nanostructured cathode surfaces have been demonstrated that couple the incident laser to resonant surface plasmon modes, which concentrate and enhance the optical-field intensity in localized areas of the surface. For example, nanogrooves designed to produce an in-groove surface plasmon resonance were patterned into gold-coated cathodes and were found to increase the photocurrent yield by 6 orders of magnitude over bare gold through multiphoton photoemission~\cite{polyakov2013nanogrooveCathodes}. Nanohole arrays designed to have a grating resonance have also been demonstrated in copper cathodes, increasing photocurrent yield 120 times over bare copper~\cite{Musumeci2013nanoholeCathode}. This approach to improving the emission characteristics is compatible with high accelerating fields. However, the photoemitted beams in these cases still had limited quality~\cite{Musumeci2013nanoholeCathode}: The optical fields were concentrated at edges, inducing emittance growth similar to the effect of surface roughness~\cite{roughness}. Also, their temporal response was not studied in detail, and may be limited by the cavity-resonance damping time. In addition, the emission is still dispersed over several micron areas in these initial demonstrations.

In this work, we propose and investigate the potential of bullseye plasmonic lenses for ultrafast, nanoscale photoemitters. Such structures concentrate optical fields to a single, central spot on a flat surface~\cite{zhang2006linPolBESNOM,chen2009radialPolarizationBE}, potentially providing aberration-free electron emission and enabling the use of nanoscale photo-triggered emitters in high-field environments.  We first show electromagnetic simulations to demonstrate how to control the spatiotemporal characteristics of the optical fields and the corresponding photoemission by geometric design. Then, we demonstrate fabrication of actual bullseye lenses in gold films by two methods and characterize their plasmonic behavior using cathodoluminescence spectromicroscopy. We show that spatial and spectral plasmonic characteristics are like predicted in simulation, supporting that ultrafast nano-emission can be achieved. Altogether, the results support the potential for bullseye lenses as high brightness electron sources and establish a new research direction in the field of plasmon-enhanced ultrafast electron nano-emission.

\begin{figure}[ht]
\includegraphics[scale=0.8]{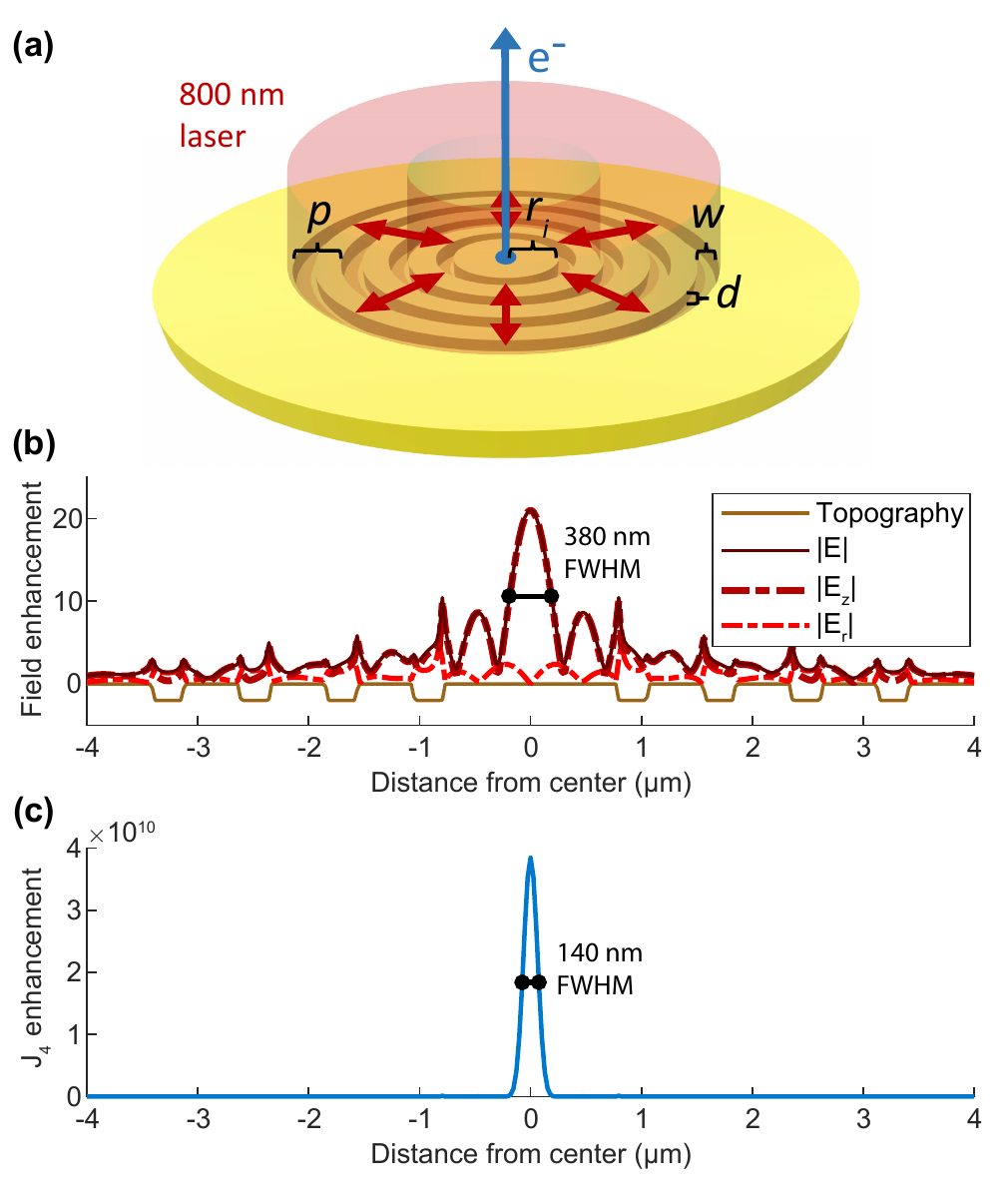}
\caption{Simulated electric field and photoemission profiles during excitation of a Au plasmonic bullseye lens with a radially polarized, 800-nm, continuous-wave (cw) laser. \textbf{a} Schematic of the bullseye geometry. The parameters shown include the grating period $p$, groove width $w$, groove depth $d$, and center plateau radius $r_{i}$. For the simulation results in following subpanels, $p$ = 783 nm, $d$ = 90 nm, $w$ = 270 nm, and $r_i$ = 783 nm. The number of rings $N$ = 4. \textbf{b} Total ($|E|$) electric field enhancement shown with its normal ($|E_z|$) and tangential ($|E_r|$) components at the bullseye surface. The field enhancement is defined relative to the peak field in the incident beam. The bullseye topography is superimposed on the plot for reference: grooves are 90 nm deep. \textbf{c} Calculated four-photon photocurrent density enhancement ($J_4$) profile at the bullseye surface using the generalized Fowler-Dubridge equation~\cite{bechtel1977generalizedFD}. FWHM is full width at half maximum.}
\label{fig:FDTDProfile}
\end{figure}

\section{\label{sec:level2} Plasmonic lens design and electromagnetic simulations}

\subsection{Lens principle and parameters}

The system under study is a nanopatterned photocathode excited by ultrafast laser pulses with an 800 nm center wavelength. The pattern consists of equally-spaced, concentric annular grooves forming a bullseye plasmonic lens. The geometry is defined by the five parameters illustrated in Figure~\ref{fig:FDTDProfile}a: The number of rings $N$, grating period $p$, groove width $w$, groove depth $d$, and center plateau radius $r_{i}$. Such gratings couple the component of incident light with electric field perpendicular to the grooves into surface plasmon polaritons (SPPs). A radially polarized laser at normal incidence is used so that the electric field direction is always perpendicular to the grooves and the launched SPPs are in phase. These SPPs then propagate and interfere to give maximum field enhancement at the structure center.

\begin{figure*}[ht!]
\includegraphics[scale=0.85]{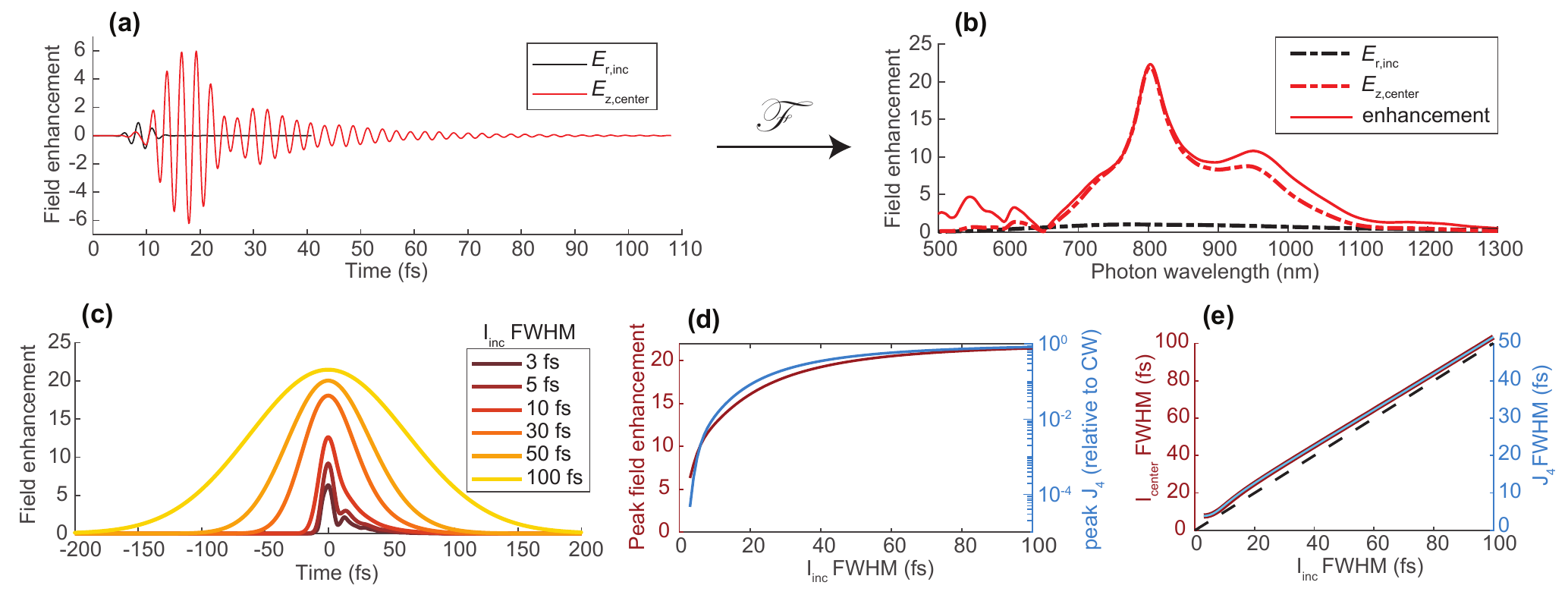}
\caption{Simulated ultrafast temporal response of the Au plasmonic bullseye lens studied in Figure~\ref{fig:FDTDProfile}. The number of rings $N$ = 4, grating period $p$ = 783 nm, groove width $w$ = 270 nm, groove depth $d$ = 90 nm, and center plateau radius $r_{i}$ = 783 nm. \textbf{a} Impulse response computed using an incident radially-polarized laser pulse with a temporal full width at half maximum (FWHM) of 3 fs. $E_{r,inc}$ is the maximum lateral electric field of the incident pulse, and $E_{z,center}$ is the normal electric field at the bullseye center. The electric field is normalized to the maximum $E_{r,inc}$. \textbf{b} Transfer function obtained by Fourier transform of the impulse response. Enhancement, ie. the magnitude of the transfer function, is shown along with the magnitudes of the Fourier transforms of $E_{r,inc}$ and $E_{z,center}$. Enhancement is obtained by dividing the Fourier transform of $E_{z,center}$ by that of $E_{r,inc}$. \textbf{c} Field enhancement temporal envelopes for varying incident pulse durations. $I_{inc}$ FWHM is the temporal full width at half maximum of the incident pulse intensity. \textbf{d} Peak field enhancement and photocurrent density ($J_4$) at the bullseye center as a function of incident pulse duration. Peak $J_4$ is plotted relative to the $J_4$ under cw illumination. The top of both y axes correspond to the value under CW illumination. \textbf{e} The temporal FWHM of the field intensity at the bullseye center ($I_{center}$) and $J_4$ as function of incident pulse duration. }
\label{fig:TROverview}
\end{figure*}

We use gold as the plasmonic material, enabling four-photon photoemission at 800 nm wavelength. Gold is oxidation resistant and provides effective photoemission surfaces; in fact, four-photon photoemission from gold cathodes patterned with linear gratings has been demonstrated~\cite{polyakov2013nanogrooveCathodes}. Gold also provides long SPP propagation lengths in the red and near-infrared range: 10-15 $\mu$m in as-deposited polycrystalline films~\cite{schuessler2009AuSPPPropLength,kuttge2008loss} and greater than 60 $\mu$m in template-stripped and single-crystal gold films~\cite{mcpeak2015plasmonic,olmon2012optical,kuttge2008loss}. 

\subsection{Nanoscale field enhancement and photoemission }
We first simulate the electromagnetic fields for a lens under continuous-wave (cw) illumination to study their spatial distribution and the expected emission spot size. An 800-nm wavelength laser is focused at the surface using a numerical aperture (NA) of 0.07, giving a donut-shaped in-plane intensity with peak-to-peak diameter of 4.5 $\mu$m. The lens has 4 rings with period $p$ of 783 nm, which is the corresponding SPP wavelength in gold. This aims to satisfy momentum conservation between the normal incidence photons and the SPPs as given by the grating equation~\cite{raetherBookSPsOnGratings}:
\begin{equation}
\mathrm{\mathbf{k}_{spp}} = \mathrm{\mathbf{k}_{photon,xy}} \pm n\mathrm{\mathbf{g}}
\label{eq:gratingEq}
\end{equation}
Here, $\mathrm{\mathbf{k}_{spp}}$ and $\mathrm{\mathbf{k}_{photon,xy}}$ are the in-plane wave vectors of the SPP and incident photon, respectively, while \textbf{g} is the grating vector and n is an integer. We set $r_i$ to be 783 nm to coincide with antinodes of the two-dimensional (2D) standing wave formed by the interfering SPPs. This causes reflections from the edge to resonate, further increasing field enhancement. With these parameters fixed, we then performed a series of finite-difference time-domain (FDTD) simulations in Lumerical~\cite{Lumerical} to optimize the depth and width of the rings for maximum field enhancement at the center. The optimum is found for $d$ = 90 nm and $w$ = 270 nm. 

The electric field magnitude profiles for the optimized structure are shown in Figure~\ref{fig:FDTDProfile}b. The field enhancement relative to the peak field of the incident laser is maximized at the center (21.0). Also, the lateral electric field $E_r$ is zero at the center and remains small relative to the normal field $E_z$ within the four-photon photoemission peak shown in Figure~\ref{fig:FDTDProfile}c. 

The anticipated four-photon photocurrent density $J_{4} \propto |\mathrm{\mathbf{E}}|^{8}$~\cite{bechtel1977generalizedFD}; the photocurrent density enhancement is shown in Figure~\ref{fig:FDTDProfile}b. Remarkably, the $J_{4}$ enhancement at the center is $3.8 \times 10^{10}$. Also, using multiphoton photoemission practically eliminates contributions from the side lobes and edges of the structure, creating a single, tightly focused emission spot with full width at half maximum (FWHM) of 140 nm in the flat center plateau.

\subsection{Ultrafast temporal response}

To study the emitter's temporal response and how it depends on the lens geometry, we performed impulse-response FDTD simulations. Using incident laser pulses with 3-fs FWHM duration, we simulate the time-resolved electric field at the structure center, $E_{\textrm{z,center}}(t)$. The spectral response in the linear intensity regime is described by the complex transfer function T($\omega$), obtained from the Fourier transforms of the time-dependent incident and enhanced fields\cite{warne2005handbook}:
\begin{equation}
E_{\textrm{z,center}}(\omega) = T(\omega)E_{\textrm{x,inc}}(\omega)
\label{eq:transferFunc}
\end{equation}
The frequency-dependent field enhancement is then given by $|T(\omega)|$. We also use eq.~\ref{eq:transferFunc} to compute the temporal response for Gaussian incident pulses of varying duration.
 
We first study the four-ring bullseye lens optimized above (see Figure~\ref{fig:FDTDProfile}). The simulated incident and enhanced electric fields over time are shown in Figure~\ref{fig:TROverview}a. Notably, the plasmonic field is stronger than the incident laser field, but lasts more than 10 fs longer. In the next section, we will show that some of this broadening is due to the delay in arrival of plasmons generated from outer rings relative to inner rings. This leads to a spectral response with finite bandwidth, shown in Figure~\ref{fig:TROverview}b as a function of incident photon wavelength. There is a strong peak at 800 nm, which we attribute to a surface plasmon resonance in the center plateau. The grooves in this structure are deep enough to disturb SPP propagation and modify their dispersion, leading to additional peaks and valleys in the response (see Figure~\ref{fig:TRDepth}). Additional peaks have been observed experimentally in transmission spectra of plasmonic lenses with a central aperture~\cite{ebbesen2001TransmissionBE} and computational work finds significant dispersion modification in deep gratings, such as opening of plasmonic bandgaps~\cite{hooper2002deepGratings}. 

Effects on the response shape for varying incident pulse length ($I_{\textrm{inc}}$ FWHM) are shown in Figure~\ref{fig:TROverview}c. Longer pulses generate a Gaussian response with peak field enhancement identical to that for cw illumination. On the other hand, for pulse lengths of tens of femtoseconds, there is substantial temporal broadening and the plasmonic field amplitude is reduced. These effects on the response strength and duration are quantified in Figure~\ref{fig:TROverview}d-e. We note that the FWHM durations in Figure~\ref{fig:TROverview}e are set by the main peak in the temporal envelope, while the tails are suppressed due to the scaling of four-photon photoemission with the intensity ($J_4(t) \propto I^4(t))$~\cite{bechtel1977generalizedFD}. As a result, the response duration is mostly linear with the input pulse duration, always about 3.5 fs longer due to the propagation delay between rings. For few-femtosecond pulse lengths, there are deviations from linearity which may be related to beating and envelope asymmetry. Still, this lens is predicted to be capable of producing sub-10 fs photoelectron pulses. 

\subsection{Tuning the response by geometric design}

\begin{figure}[ht]
    \centering
    \includegraphics[scale=0.8]{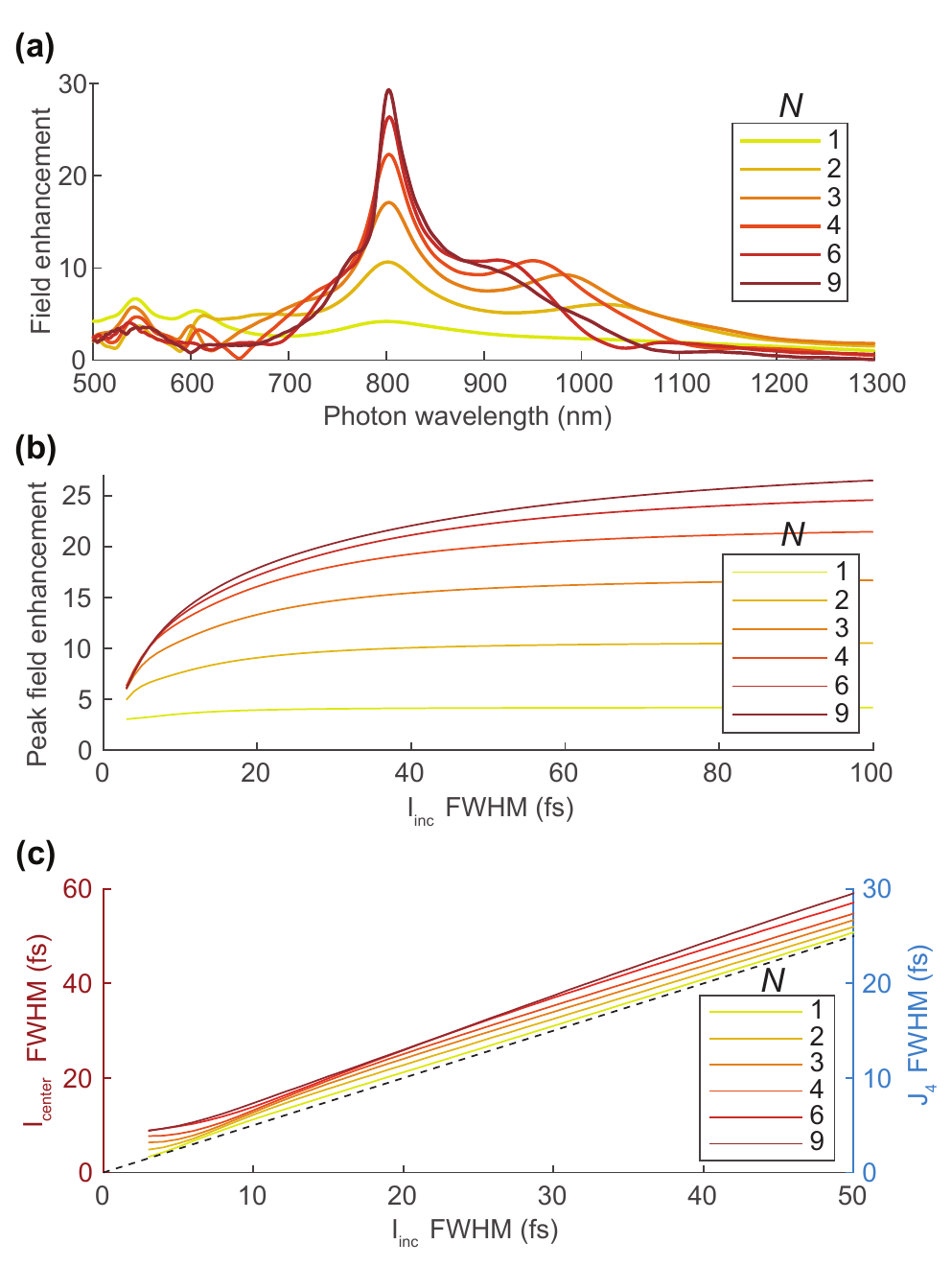}
    \caption{Simulated ultrafast temporal response of bullseye lenses with varying number of rings, $N$. The grating period $p$ = 783 nm, groove width $w$ = 270 nm, groove depth $d$ = 90 nm, and center plateau radius $r_{i}$ = 783 nm. \textbf{a} Transfer function computed for lenses with $N$ ranging from 1 to 9. \textbf{b} Maximum field enhancement at the lens center for incident pulses with varying temporal full width at half maximum (FWHM). \textbf{c} Temporal FWHM of the electric field intensity at the lens center ($I_{\textrm{center}}$) for incident pulses with varying temporal FWHM. The temporal FWHM of the four-photon photocurrent density $J_4$ is half of the $I_{\textrm{center}}$ FWHM.}
    \label{fig:TRNRings}
\end{figure}

We now present a series of impulse-response simulations while varying geometric parameters to clarify design rules for ultrafast photoemission applications. We first vary the number of rings, $N$, obtaining the simulated spectral response shown in Figure~\ref{fig:TRNRings}a. For $N$ = 1, the bandwidth extends over the entire wavelength range studied, and there is a plasmonic resonance peak at 800 nm. Adding more rings increases the field enhancement by coupling more light, but it shrinks the bandwidth by increasing the SPP propagation distance, and hence delay time, between inner and outer rings. This generally increases the peak field enhancement for varying incident pulse length shown in Figure~\ref{fig:TRNRings}b while increasing the response duration as shown in Figure~\ref{fig:TRNRings}c. For few-femtosecond pulses, however, there is a limit to the number of added rings that increase field enhancement, beyond which the delay time between inner and outer plasmons is too long for them to overlap. In this limit, adding additional rings only increases the pulse duration without increasing field enhancement, shown in Figure~\ref{fig:TRNRings}b-c at the shortest pulse durations.

\begin{figure}[ht]
    \centering
    \includegraphics[scale=0.8]{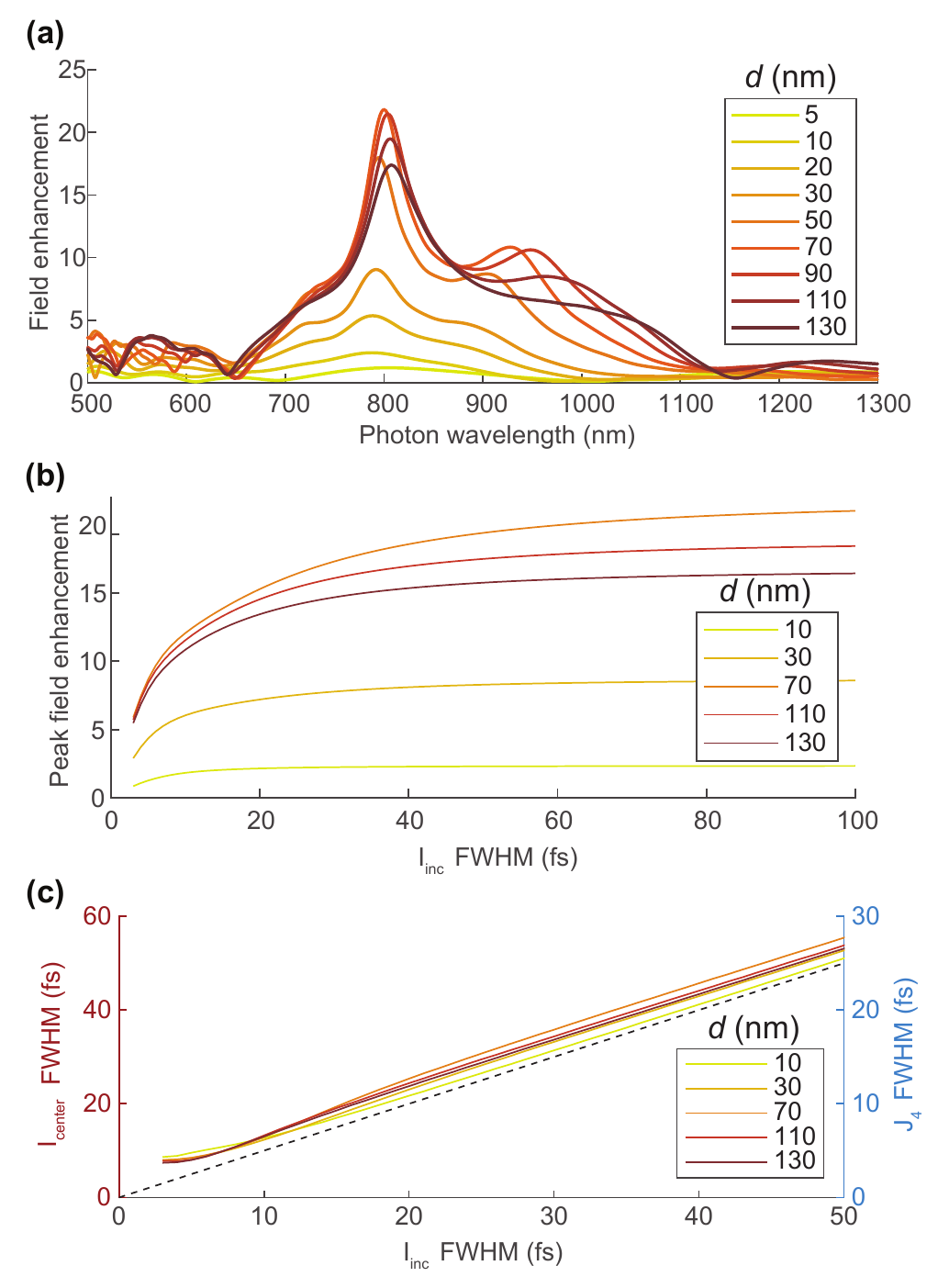}
    \caption{Simulated ultrafast temporal response of bullseye lenses with varying groove depth, $d$. The number of rings $N$ = 4, grating period $p$ = 783 nm, groove width $w$ = 270 nm, and center plateau radius $r_{i}$ = 783 nm. \textbf{a} Transfer function computed for lenses with varying $d$. \textbf{b} Maximum field enhancement at the lens center for incident pulses with varying temporal full width at half maximum (FWHM). \textbf{c} Temporal FWHM of the electric field intensity at the lens center ($I_{\textrm{center}}$) for incident pulses with varying temporal FWHM. The temporal FWHM of the four-photon photocurrent density $J_4$ is half of the $I_{\textrm{center}}$ FWHM.}
    \label{fig:TRDepth}
\end{figure}

\begin{figure}[ht]
    \centering
    \includegraphics[scale=0.8]{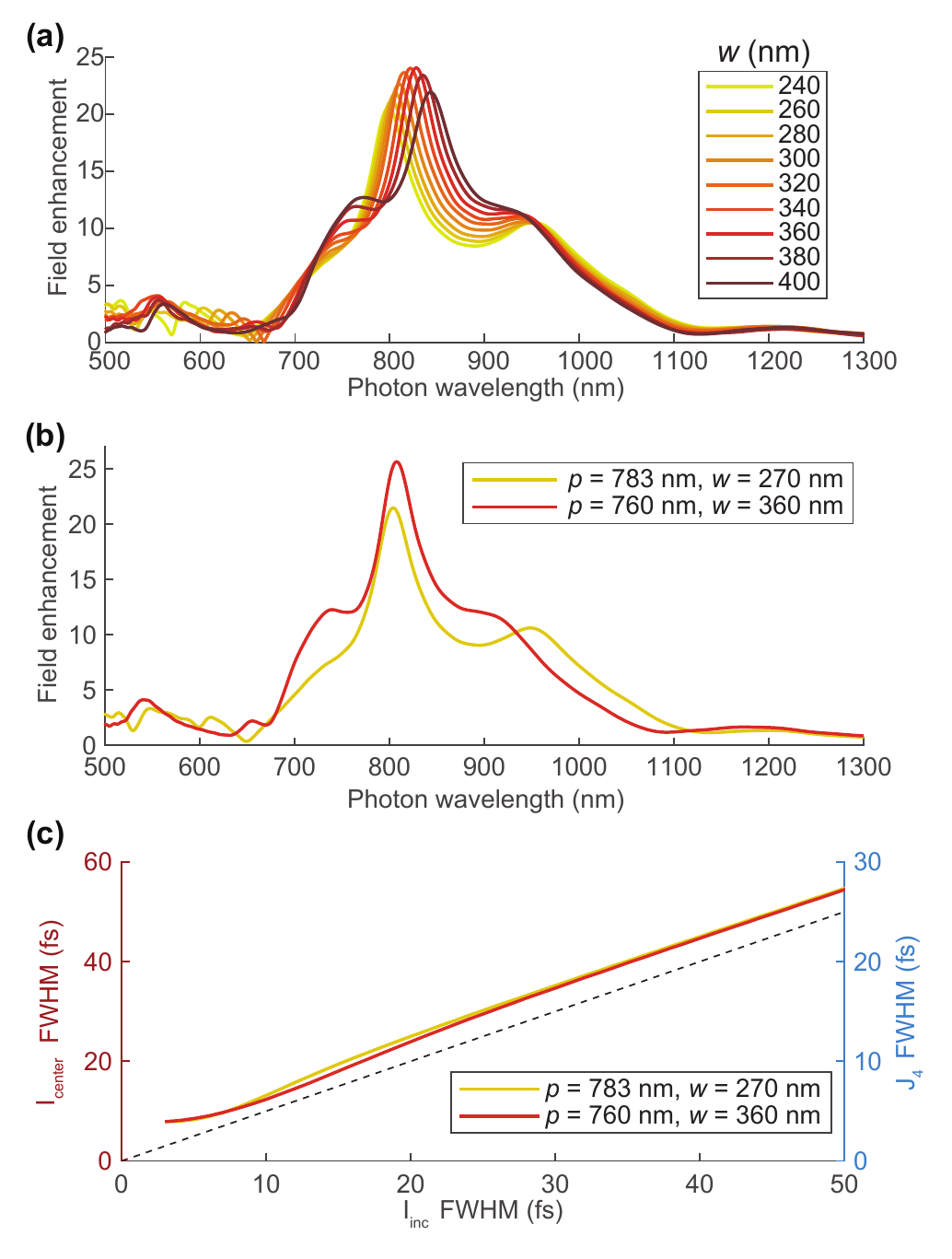}
    \caption{Simulated ultrafast temporal response of bullseye lenses with varying groove width, $w$. The number of rings $N$ = 4, groove depth $d$ = 90 nm, and center plateau radius $r_{i}$ = 783 nm. \textbf{a} Transfer function computed for lenses with varying $w$ and grating period $p$ = 783 nm. \textbf{b} Transfer function computed for the lens studied in Figure~\ref{fig:FDTDProfile} ($p$ = 783 nm, $w$ = 270 nm) and for a lens with $w$ chosen to maximize field enhancement and $p$ adjusted to center the peak wavelength near 800 nm ($p$ = 760 nm, $w$ = 360 nm). \textbf{c} Response duration for the two lenses in b. This is plotted as temporal FWHM of the electric field intensity at the lens center ($I_{\textrm{center}}$) for incident pulses with varying temporal FWHM. The temporal FWHM of the four-photon photocurrent density $J_4$ is half of the $I_{\textrm{center}}$ FWHM.}
    \label{fig:TRWidth}
\end{figure}

We then vary the depth of the grooves, $d$, obtaining the spectral response shown in Figure~\ref{fig:TRDepth}a. As $d$ increases, the field enhancement increases, saturates, and eventually decreases. While deeper grooves couple more incident photons to SPPs, they also inhibit SPP propagation to the center, leading to an optimal depth that maximizes field enhancement. The resonance peak at 800 nm emerges and grows with increasing $d$ as reflectivity of plasmons from the grooves increases. Other peaks and valleys also emerge, suggesting the deeper grooves are disturbing plasmon propagation and modifying their dispersion and interference. The peak field enhancement for varying incident pulse length scales similarly with $d$ as shown in Figure~\ref{fig:TRDepth}b. Again for few-femtosecond pulses, plasmons from the outer rings cannot reach those from the inner rings and so the field enhancement is reduced. 
There is also a noticeable effect of $d$ on the temporal response duration as shown in Figure~\ref{fig:TRDepth}c. For longer pulses, the duration seems to scale with the strength of the resonance contribution, which would extend the plasmonic field duration. Overall, $d$ has less of an effect on the pulse duration than $N$.

Next, we vary the ring width, $w$, computing the transfer functions shown in Figure~\ref{fig:TRWidth}a. Maximal field enhancement is achieved for $w$ = 360 nm, close to half the period. Adjusting the width also shifts the resonance peak. Wider groove obstacles may more strongly obstruct plasmon propagation, affecting the plasmon dispersion relation. This shift in resonance peak can be compensated by adjusting the bullseye grating period, $p$. The transfer function and temporal response of a four-ring lens optimized by allowing variable $p$ is shown with the one obtained by fixing $p$ = 783 nm in Figure~\ref{fig:TRWidth}b. The peak field enhancement and the symmetry of the transfer function are improved by using $w$ = 360 nm and compensating for the resonance peak shift by setting $p$ = 760 nm. This provides a factor-of-4 increase in the estimated four-photon photoemission yield. This comes without cost in response duration, as shown in Figure~\ref{fig:TRWidth}c. 

These results lead to a few design rules for ultrafast applications. The depth and width of the grooves should usually be optimized for maximum field enhancement. The period can then be adjusted to center the resonance peak at the desired wavelength. Finally, the structure should have as many rings as possible to maximize photocurrent while maintaining enough bandwidth to achieve the required pulse duration. The optimal geometry ultimately depends on the photocurrent and pulse duration required.

\section{\label{sec:level3}Fabrication}

\begin{figure}[ht]
\includegraphics[scale=1]{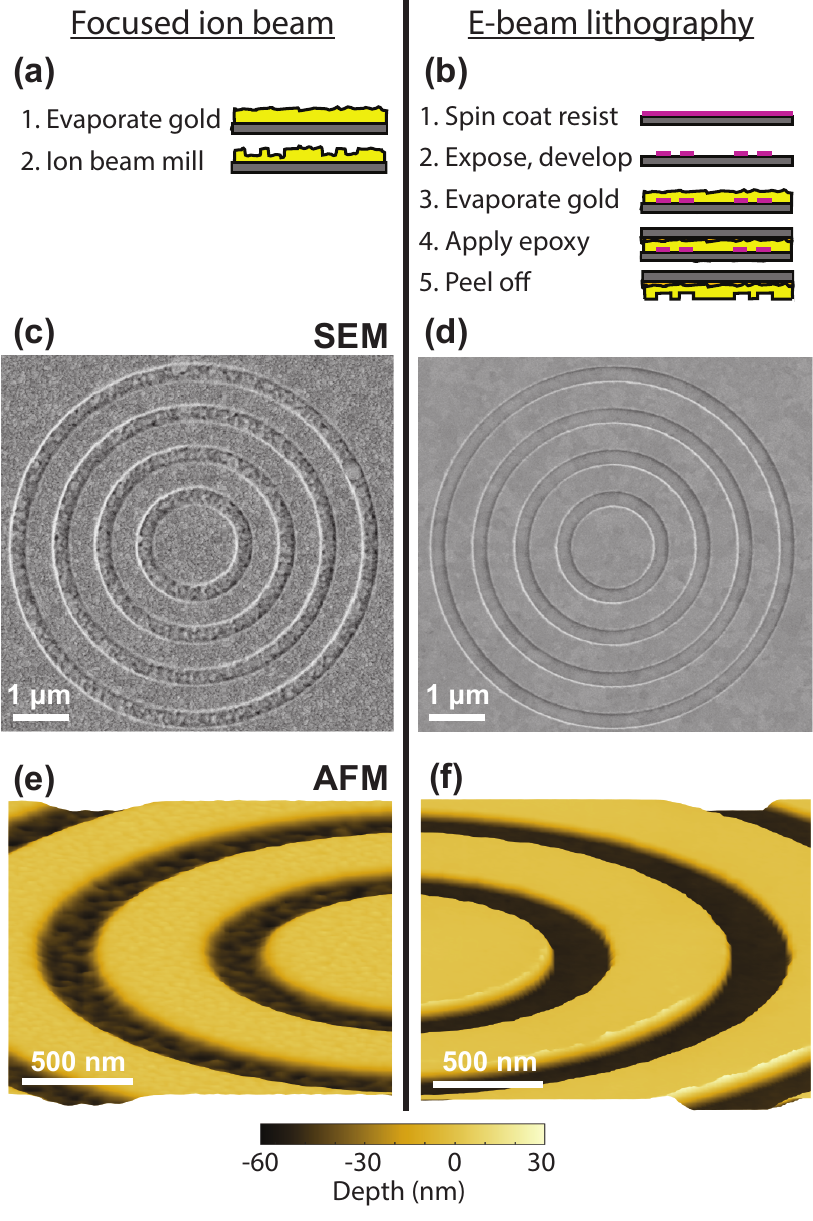}
\caption{Bullseye lenses fabricated by two methods. Process steps shown for \textbf{a} FIB milling and \textbf{b} EBL with template stripping. In-lens SEM images shown of the bullseye lenses made using \textbf{c} FIB and \textbf{d} EBL. AFM surface topography maps shown of lenses made using \textbf{e} FIB and \textbf{f} EBL. Maps are displayed to-scale in 3D at a 40$^{\circ}$ tilt.}
\label{fig:Fabrication}
\end{figure}

We fabricated bullseye lenses using two processes illustrated in Figure~\ref{fig:Fabrication}. One process involves thermally evaporating 5 nm of titanium as an adhesion layer and 150 nm of gold onto a Si wafer, then carving out the rings using focused-ion-beam (FIB) milling. A Zeiss Crossbeam 1540 EsB is used for the FIB. The other process uses electron-beam lithography followed by template stripping to produce high-precision, smooth cathode surfaces~\cite{vogel2012flat}. A negative e-beam resist hydrogen silsesquioxane (HSQ) $2\%$  is spun at 1000 rpm onto a Si wafer. The resist is then exposed using a Vistec VB300 electron-beam lithography (EBL) system and developed, leaving the designed pattern in the form of amorphous silica on the wafer. Then, 150 nm of gold was deposited onto the template, entirely covering it. Finally, the patterned gold is peeled off using an electrically and thermally conductive ultra-high-vacuum compatible epoxy resin. Atomic force microscopy (AFM) measurements confirm that similar groove depths can be made in both structures, but the EBL and template stripping process yields smoother surfaces and grooves. In fact, structures made by EBL have nearly atomically flat central areas, which minimizes degradation of the emitted electron beam from surface roughness and imperfections~\cite{roughness} (see Table~\ref{tableAFM}). In these lenses, the gratings are 50 nm deep with parameters otherwise matching the first case (783-nm period, 270-nm width, four rings): its simulated spectral response is shown in Figure~\ref{fig:TRDepth}.

\begin{table}[ht]
\caption{Topographic comparison between EBL and FIB bullseye lenses via AFM}
\label{tableAFM}
\begin{tabular}{l c c}
\hline
& FIB & EBL \\
\hline
Groove depth (nm) & 51 & 54\\
Center RMS roughness (nm) & 1.4 & 0.4\\
Groove RMS roughness (nm) & 3.2 & 0.6\\
\hline
\end{tabular}
\end{table}

\section{Cathodoluminescence spectromicroscopy}

\subsection{Examining plasmonic properties}

Here, we present cathodoluminescence (CL) spectromicroscopy measurements of plasmonic characteristics of fabricated lenses. In CL, an electron beam is focused on the sample and, by one or more mechanisms, light is emitted from the material~\cite{deAbajoOpticalExcitationsReview}. For plasmonic structures, the relevant mechanism is the broadband generation of SPPs by the fast-moving electrons as they strike the material surface. The SPPs propagate radially outward from a nanometric spot, which can then couple out to light through the bullseye grating ~\cite{polmanCLLinearGrating,atwaterBullseyeCL}. When the electron beam is at the bullseye center, circularly symmetric SPPs are excited like those that would be generated by a radially polarized laser.

We use CL to measure a few key plasmonic properties of these structures. For one, the plasmonic resonance in the central region can be mapped using CL spectromicroscopy. CL intensity has been linked to the radiative local density of optical states (LDOS) at the electron beam position, ie. the number of optical modes available at the excitation position that produce light~\cite{polmanCLLinearGrating,kociak2015LDOS}. The LDOS is enhanced by plasmonic resonance, so CL can be used to spatially and spectrally resolve resonance modes~\cite{polmanCLReview,kociakNanoOpticsEMReview}. Spatial homogeneity, circular symmetry, and a strong, sharp central peak are desirable for photoemission applications. The center wavelength and bandwidth of this resonance can be extracted from the CL spectrum obtained at the structure center. Also, the radial extent and circular symmetry of plasmon propagation and grating coupling can be inferred by angle-resolved imaging of the CL far field, which is essentially a Fourier transform of the real-space emission profile of the structure. These are important to characterize since significant propagation losses or asymmetries in coupling would reduce the field enhancement under laser illumination.

We use a modified Zeiss Gemini SUPRA 55 SEM for our CL measurements. The sample is positioned at the focal point of a horizontal Al parabolic mirror with 1 mm focal length. A 10-keV electron beam is focused onto the sample, and the emitted light is collected by the parabolic mirror over a wide angle range ($\textrm{0}^{\circ}$ - $\textrm{80}^{\circ}$ from normal) and over the entire visible spectrum and beyond. The sample was tilted by about $\textrm{25}^{\circ}$ so that the highly directed, normal emission would not escape through the entry hole in the mirror for the electron beam. We do not expect this tilt to change the plasmonic response; there will not be significant arrival time delay across the electron beam since it is much smaller than the plasmon wavelength. Also, the electron beam only generates SPPs at the surface, so the source size is only made about 10\% larger along the tilt direction.

\begin{figure*}[ht]
\includegraphics[scale=0.83]{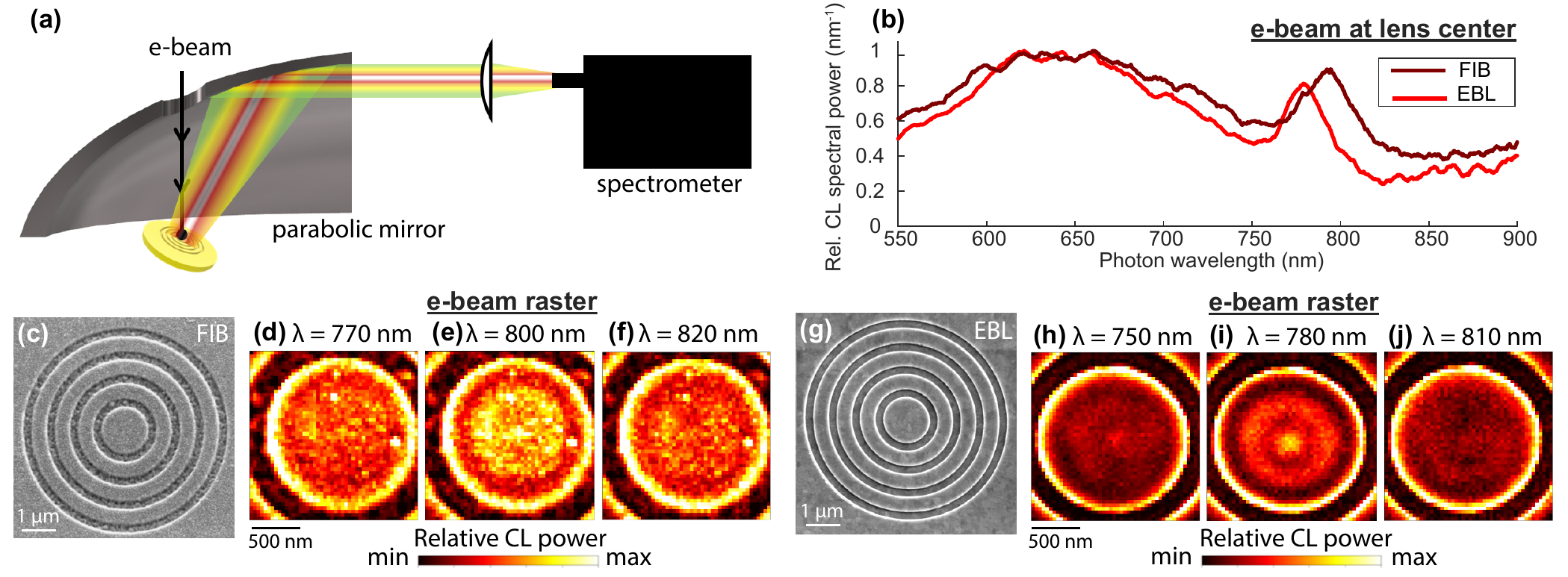}
\caption{Cathodoluminescence (CL) spectromicroscopy of the plasmonic resonance. \textbf{a} A schematic of the CL spectromicroscopy technique. The electron-induced multicolor luminescence from a tilted lens is collected by a parabolic mirror and focused by a lens into a spectrometer. The electron beam is scanned and a full CL spectrum is collected at each beam position. \textbf{b} The average of CL spectra collected within 75 nm of the structure center, shown for one lens made by FIB milling and one by EBL. \textbf{c} A SEM image and \textbf{d-f} CL spatial maps of the lens made by FIB. \textbf{g} SEM image and \textbf{h-j} CL spatial maps of the lens made by EBL. CL spatial maps are obtained from the spectromicroscopy dataset by integrating over Gaussian wavelength bands with 2$\sigma$ = 10 nm. Each map is labeled with their corresponding center wavelength, $\lambda$.}
\label{fig:ScanningCL}
\end{figure*}

\subsection{CL spectromicroscopy of plasmonic resonance}

We use scanning CL spectromicroscopy to probe the spectral and spatial characteristics of the surface-plasmon resonance in our fabricated lenses. The electron beam is rastered step by step over the central plateau of the structure, and a full CL-emission spectrum is collected at each beam position. Spectra are obtained by focusing the reflected light from the mirror onto a multi-mode optical fiber with a 200 $\mu$m diameter and then dispersing it using a spectrometer consisting of an Acton 2300i monochromator (150 line/mm, 500 nm blazed grating) and Andor Newton electron-multiplied charge-coupled device (CCD). The dark current background is subtracted and spectra are normalized by the instrument response over the measured wavelength range. The open-source Python-based ScopeFoundry software developed to control this experiment is available online for further reference~\cite{durhamScopeFoundry,ScopeFoundryWebsite}. 

From this 3D data set, we can extract average emission spectra from regions of interest. Average spectra over the region within 75 nm of the bullseye center are shown in Figure~\ref{fig:ScanningCL}a. Spectra shown here are smoothed using a second-order Savitsky-Golay filter with 15-nm window. The CL emission spans the entire detection spectral range. This is because a SPP of any wavelength from the broadband range generated by the electron beam can satisfy the grating equation and couple to light at a wavelength-dependent emission angle, and nearly all emission angles are collected by the parabolic mirror. However, there is a notable emission peak near 800 nm where a surface plasmon resonance is expected. 

We can then examine the spatial profile of the plasmonic resonance. We apply virtual Gaussian bandpass filters with 2$\sigma$ = 10 nm to the entire dataset, yielding maps of CL emission over narrow spectral bands as a function of beam position [see Figure~\ref{fig:ScanningCL}c-j]. For maps of emission near the resonance wavelength, a zero-order Bessel-function spatial profile is observed corresponding to a cylindrical plasmonic resonance of the central plateau. For the maps at wavelengths just outside of the resonance peak, such a spatial profile is not observed, confirming that it is a resonance effect. For the structure made by EBL, the CL intensity is higher when the structure is excited at the central peak antinode than at the nearby annular antinode. For the structure made by FIB, however, the CL intensity is similar when exciting at either antinode. This suggests that the constructive interference of SPP modes from different directions is improved in the EBL structures, leading to a stronger central peak.

Other features in the maps are also present off resonance. For instance, bright spots are present in the map of the FIB-milled structure [Figure~\ref{fig:ScanningCL}d-f] which correspond to bright signals in the secondary electron image [Figure~\ref{fig:ScanningCL}c]. The groove edges are brighter in both structures regardless of wavelength. These correspond to topographic features with locally high surface area and roughness, which can enhance the radiative LDOS by scattering more SPP modes out to light. The near-atomic smoothness of the central region in the EBL structures eliminates the scattering sites observed in the FIB structure. This will reduce damping of the resonance and thereby improve the field enhancement under laser illumination and consequently the amount of emitted electrons.

We note that there are key differences between CL mapping and the FDTD simulations used in section II. The excitation mechanism is different: FDTD simulates a radially polarized laser excitation and calculates the field enhancement at all positions, whereas CL mapping rasters an e-beam and collects the integrated light output as a function of e-beam position. This scanning local excitation provides a different spatial profile than expected for a global laser excitation based on FDTD. For instance, the CL emission is stronger when the electron beam is positioned at edges than the structure center, but this is not expected for the case of radially polarized illumination. Still, we can use CL mapping to visualize scattering sites and study the symmetry and smoothness of the plasmonic resonance mode, as discussed above. For more detailed discussion of the CL dependence on electron beam position, see the Appendix.

\begin{figure}[ht]
\includegraphics[scale=0.8]{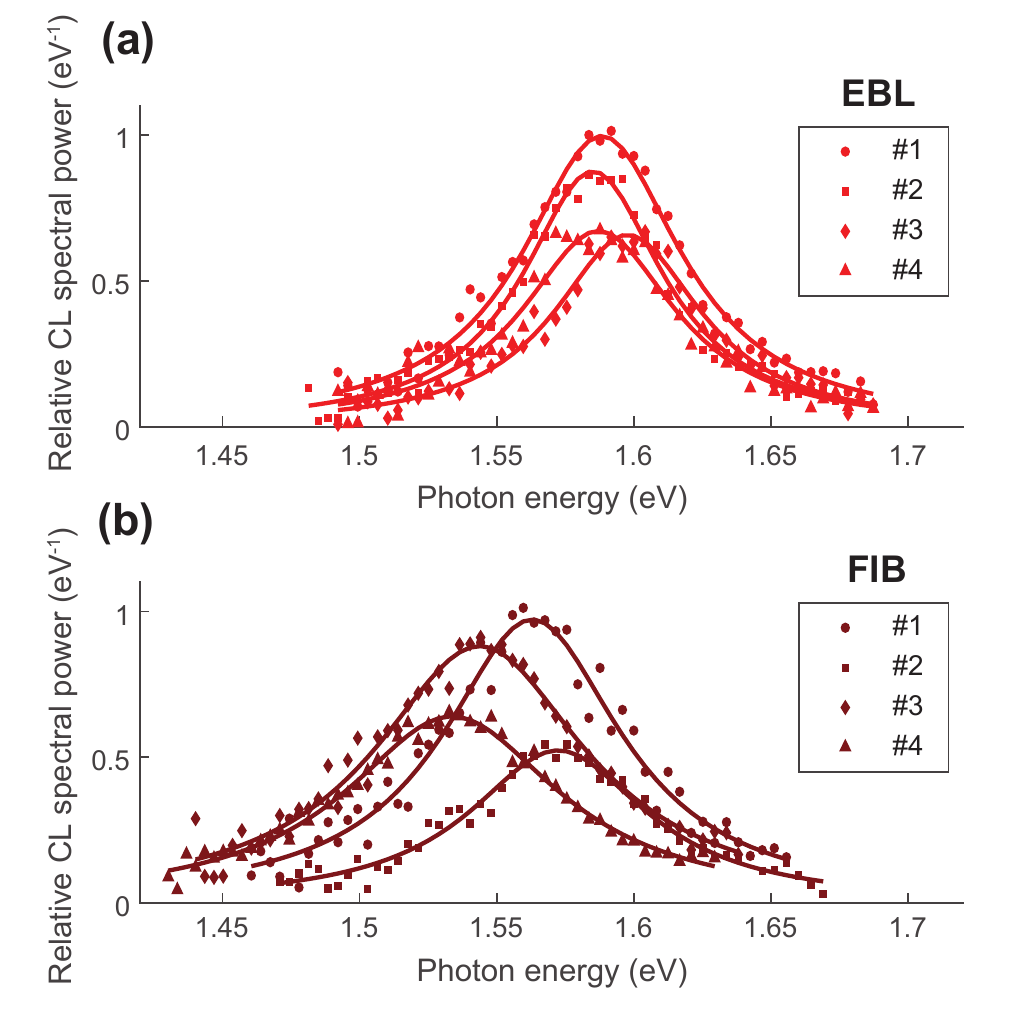}
\caption{A comparison of the plasmonic-resonance characteristics of bullseye lenses made by EBL and FIB milling using CL spectroscopy. Resonance peaks are shown for four lenses made using \textbf{a} EBL and \textbf{b} FIB. The spectra are fitted with a parabolic background plus a Lorentzian over a 0.2-eV range centered at the peak. Points indicate data over the fitting window after the background is subtracted, while lines indicate the Lorentzian peak fit to the data.}
\label{fig:CLPeakFitting}
\end{figure}

To quantify and compare the resonance characteristics of structures made by FIB and EBL, we fit the resonance peaks in CL spectra collected using an e-beam at the lens center (where the plasmons generated match the circular symmetry of the structure) to determine the resonance wavelength and the FWHM. Average spectra from scan positions within 75 nm of the bullseye center are extracted from spectromicroscopy scans. The spectra are then re-binned into 2-nm wavelength bins and converted from wavelength to energy scale. The resonance peak is fit using a Lorentzian plus a parabolic background. Four structures made by each fabrication method were measured and analyzed. The converted spectra and peak fitting are shown in Figure~\ref{fig:CLPeakFitting}. The average resonance wavelength is 798.4 $\pm$ 8.8 nm standard deviation for the FIB-milled structures and 779.6 $\pm$ 2.4 nm for the EBL structures. Both are near the 794-nm resonance wavelength predicted by FDTD for structures with the fabricated dimensions. The average measured frequency-to-FWHM ratio, or Q factor, is 17.8 $\pm$ 2.0 for the FIB-milled structures and 23.9 $\pm$ 1.9 for the EBL structures. The more precise resonance wavelength and higher Q factor (lower damping) of the EBL structures can be attributed to the reproducibility of the structure dimensions and their smoothness. 

\subsection{Angle-resolved CL and photon-plasmon coupling}

We study the angle-resolved CL emission to infer the radial extent of effective plasmon propagation and grating coupling in the fabricated lenses. We both simulate and measure the CL far field for four- and 12-ring bullseye lenses. The far field is measured by positioning the electron beam at the structure center and imaging the parabolic mirror~\cite{coenenAPLAngleResolvedCL,coenenBookAngleResolvedCL}. Because the emitting structure (several micrometers) is small compared to the distance from the mirror (1 mm), the position where an emitted photon reflects from the mirror is determined by the emission angle. The parabolic mirror brings the emission to infinity focus, which is filtered by a bandpass filter centered at 800 nm with 40 nm bandwidth and then magnified and imaged onto a ThorLabs DCC3260M CMOS camera. For each measurement, 30 one-second exposures of the CCD are acquired and averaged. A background image with the beam blanked is acquired under the same conditions and is subtracted from the beam-on image. Then, each pixel in the image is mapped to an emission angle, and the signal is normalized by the solid angle collected by that pixel to give an intensity map. The measured far field is then corrected for tilt and rotation of the sample relative to the mirror. 

Using FDTD, the CL process can be numerically modeled~\cite{Lumerical}. An impinging 10-keV electron at the center of the bullseye is modeled using a series of dipoles normal to the surface. The dipoles are delayed in phase to create a propagating, localized source of electric field. This generates a time- and z-dependent current density close to that of a moving electron~\cite{chaturvedi2009CLDipoleFDTD}: $\vec{J}(t,z)= -ev\hat{z} \delta(z-vt)\delta(x-x_{0})\delta(y-y_{0})$. Here, $e$ is the fundamental charge, $v$ is the speed of the impinging electron, $\hat{z}$ is the unit vector in the z direction, and $x_0$ and $y_0$ give the lateral position of the impinging electron. The resulting field decays laterally, and vanishes within a few nm in the metal; therefore, no field can directly couple to the grooves, and all observed emission is due to generation, propagation, and outcoupling of SPPs. To avoid abrupt appearance and disappearance of the dipole field, which would create stray fields, a raised-cosine filter is used to gradually increase and decrease the amplitude oscillations of the starting and ending dipoles in time. The simulation box was gradually increased in order to collect larger angles and mesh dimension was decreased to achieve convergence.

\begin{figure}
\includegraphics[scale=0.8]{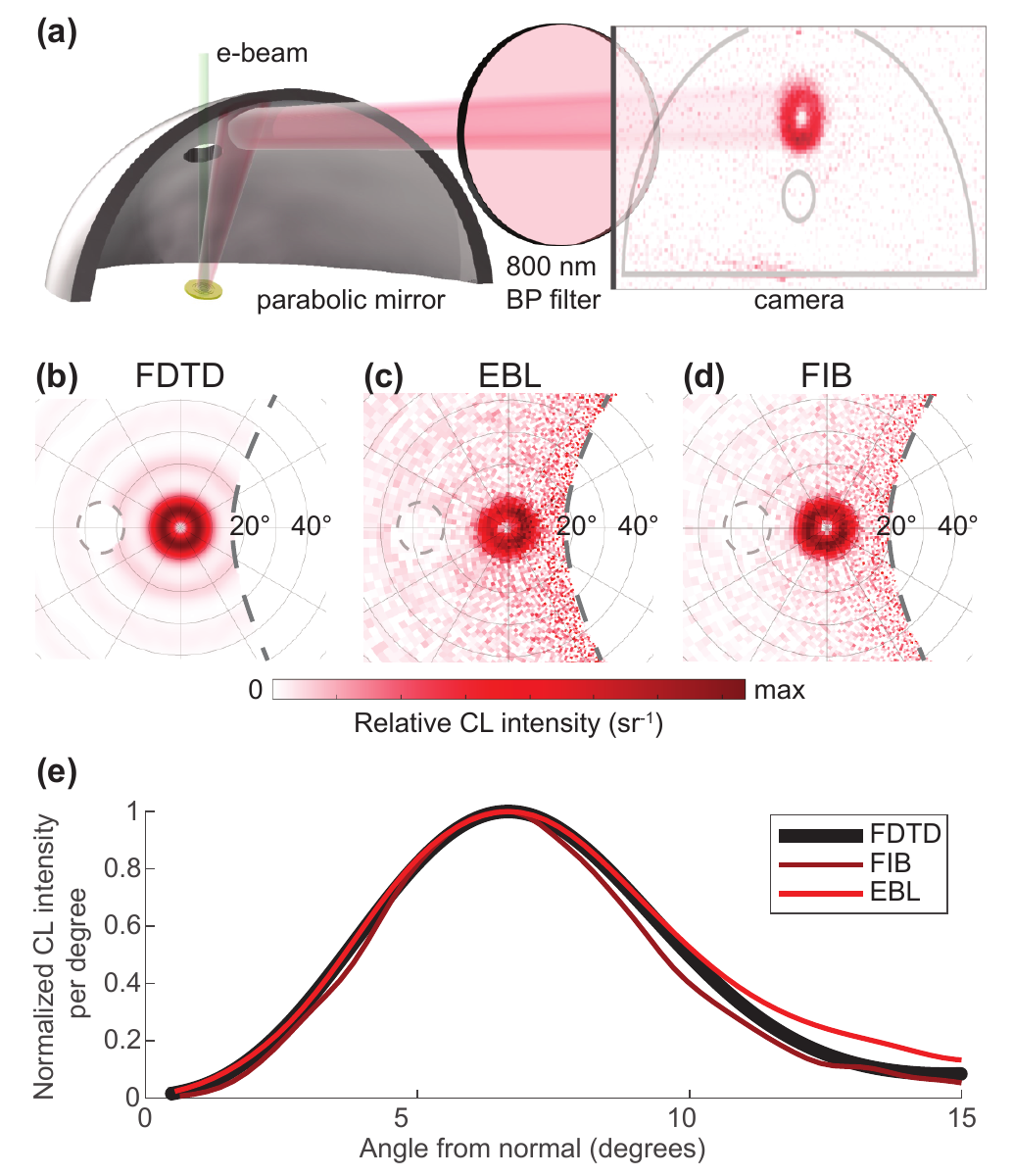}
\caption{Angle-resolved cathodoluminescence (CL) with e-beam at the structure center. \textbf{a} A schematic of the Fourier imaging technique. The parabolic mirror is imaged through a band-pass (BP) filter centered at 800-nm wavelength with 40-nm bandwidth. Far field plots are then generated by transforming mirror coordinates to angular coordinates. The sample is tilted at 25$^{\circ}$ so emission is not lost through the entry hole for the electron beam. \textbf{b} The FDTD-simulated far field. CL far field polar plots for lenses made using \textbf{c} EBL and \textbf{d} FIB. The gray dashed lines trace the electron-beam entry hole and the open face of the parabolic mirror. \textbf{e} CL distributions per degree from normal obtained by azimuthal integration of the far field.}
\label{fig:CLFarField}
\end{figure}

Simulated and measured far-field CL of four-ring lenses are shown in Figure~\ref{fig:CLFarField}a-c. A donut beam is observed in all cases, supporting that the emission is radially polarized. Other work performing CL polarimetry on bullseyes has resolved the angle-dependent polarization state and verified that the emission is radially polarized~\cite{polmanBullseyeCLPolarimetry}. Both the angle of maximum emission and the overall angular distribution measured for lenses made by EBL and FIB closely match that predicted using FDTD (see Figure~\ref{fig:CLFarField}d). This indicates that the radial distribution of CL emission from plasmon-photon coupling is robust to fabrication imperfections, including the surface roughness of the FIB-milled structure. These measurements also show that this normally directed emission is captured effectively at 25$^{\circ}$ sample tilt; this supports the validity of the spectra shown above and demonstrates the need for tilting the sample to study this emission.

\begin{figure}
\includegraphics[scale=0.80]{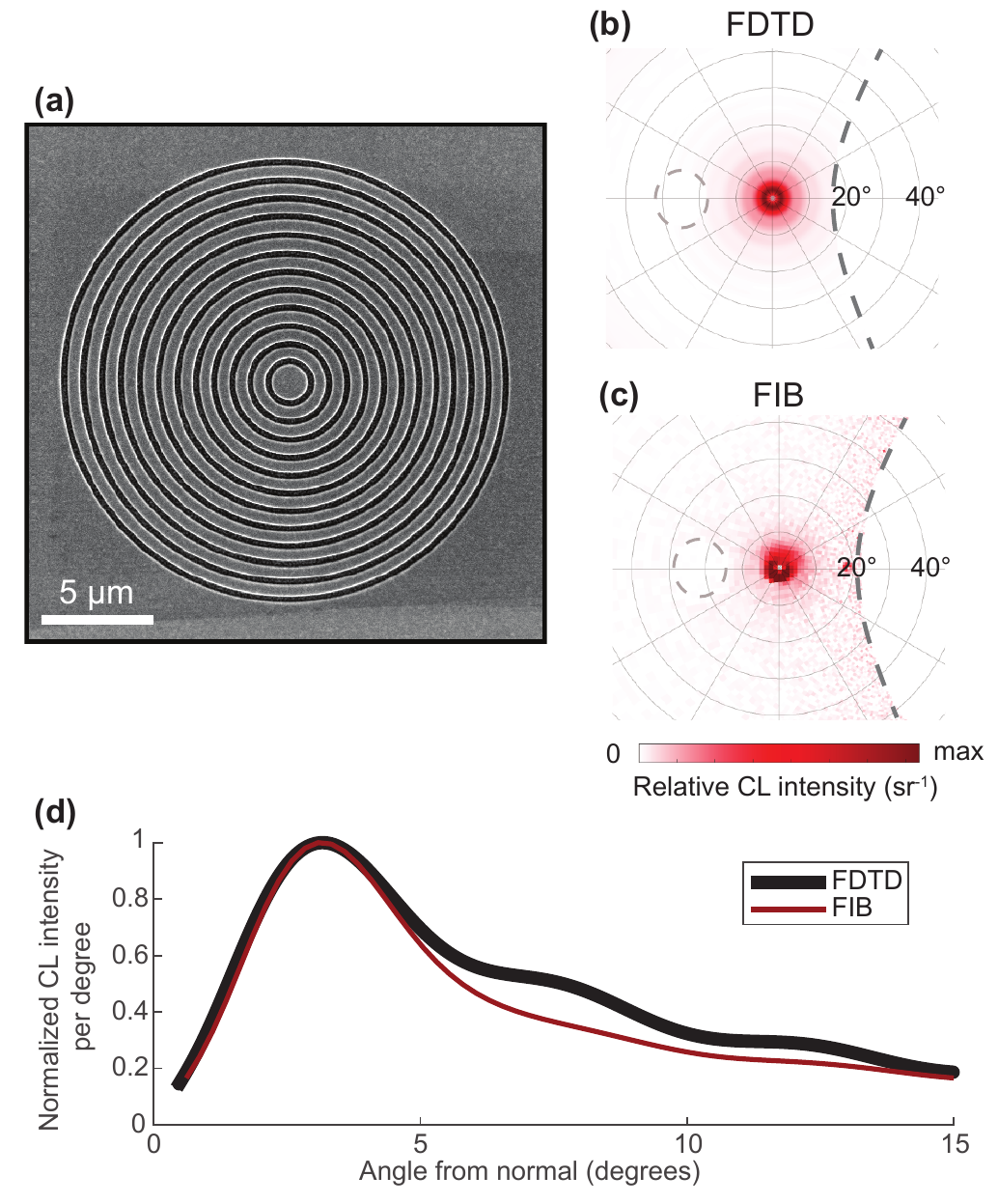}
\caption{Angle-resolved cathodoluminescence (CL) for a 12-ring lens milled using focused ion beam (FIB). \textbf{a} A SEM image of the 12-ring lens. The FDTD-simulated (\textbf{b}) and experimentally measured (\textbf{c}) CL far-field polar maps for the electron beam positioned at the center. The gray dashed lines trace the electron-beam entry hole and the open face of the parabolic mirror. \textbf{d} The emission distribution along angle from normal obtained by azimuthal integration of the far field.}
\label{fig:CLFarField12Ring}
\end{figure}

We also fabricated and studied a 12-ring bullseye lens using FIB as shown in Figure~\ref{fig:CLFarField12Ring}a. The simulated and measured far-field for this structure are shown in ~\ref{fig:CLFarField12Ring}b-c. Again, a donut beam is produced as expected. The far field is more asymmetric than for the four-ring structures, which may be due to challenges in making the outer rings accurately concentric with the inner rings using FIB. Still, the angular breadth of emission is greatly reduced, indicating that the plasmons can outcouple over a greater radial extent. As highlighted in Figure~\ref{fig:CLFarField12Ring}d, the peak emission angle is 3$^{\circ}$, whereas in the four-ring lenses it was 7$^{\circ}$. The distribution matches well between simulation and experiment overall, especially at less than 5$^{\circ}$. This suggests that additional plasmon propagation losses due to polycrystallinity are not a limiting factor even for this 12-ring structure. 

\section{Conclusions}

We have presented a novel design for ultrafast nanoscale electron emitters which has several advantages for the next generation of electron-based instrumentation. First, the design is compatible with high accelerating fields while decreasing the source size by two orders of magnitude as compared to an unpatterned flat cathode. Electromagnetic simulations and cathodoluminescence mapping support that plasmonic resonance leads to a single dominant central peak in optical intensity which could be used to generate electron pulses with 140-nm FWHM (60-nm RMS) lateral size. Second, the emission surface can be made nearly atomically flat, suppressing emittance increase from surface roughness and from aberrations due to field curvature, differently from what happens with tips. Thirdly, the geometric parameters can be tuned to optimize the photocurrent and temporal response for application requirements without changing the emission spot size. The computations and experiments shown here support that the plasmonic properties and the spectral bandwidth of such structures are compatible with emission of sub-10-fs pulses. 

These plasmonic lenses could also facilitate emerging high-intensity modes of operation. For instance, they reduce the laser power required to access the optical field emission regime, in which the fields are strong enough to modulate the work function at the optical frequency. This operating regime is of great interest because it allows control of the photocurrent density at attosecond timescales, providing the potential to generate attosecond electron-pulse trains~\cite{hommelhoff2012opticalFieldEmission}. In addition, the compatibility of these lenses with few-cycle pulses could allow control of the photocurrent intensity by tuning the carrier envelope phase, as has been demonstrated for tip emitters~\cite{lienau2014cepPhotoemission}.

In terms of applications, the transverse brightness of such nanocathodes is expected to be more than 1 order of magnitude better than present state of art flat cathodes. Their use in rf environments would enable production of relativistically accelerated electrons with picometer emittance, which could be focused down to nanometer sizes and efficiently injected into advanced acceleration devices~\cite{DLA}. In ultrafast electron imaging, such sources promise to bridge the gap in spatial resolution between static and ultrafast relativistic sources, applying electron-based characterization to nanoscale dynamics.

\begin{acknowledgements}
D.F. and the work at the Molecular Foundry were supported by the Office of Science, Office of Basic Energy Sciences, of the U.S. Department of Energy under Contract No. DE-AC02-05CH11231. Funding for D.B.D. was provided by STROBE: A National Science Foundation Science and Technology Center under Grant No. DMR 1548924. We thank F. Ogletree, S. Aloni, and E.S. Barnard at the Molecular Foundry for their advice and assistance with the CL setup. We thank S. Dhuey at the Molecular Foundry for assistance with the e-beam lithography.

D.B.D. and F.R. contributed equally to this work.

\end{acknowledgements}

\appendix*
\section{CL dependence on electron beam position}

\begin{figure}
\includegraphics[scale=0.80]{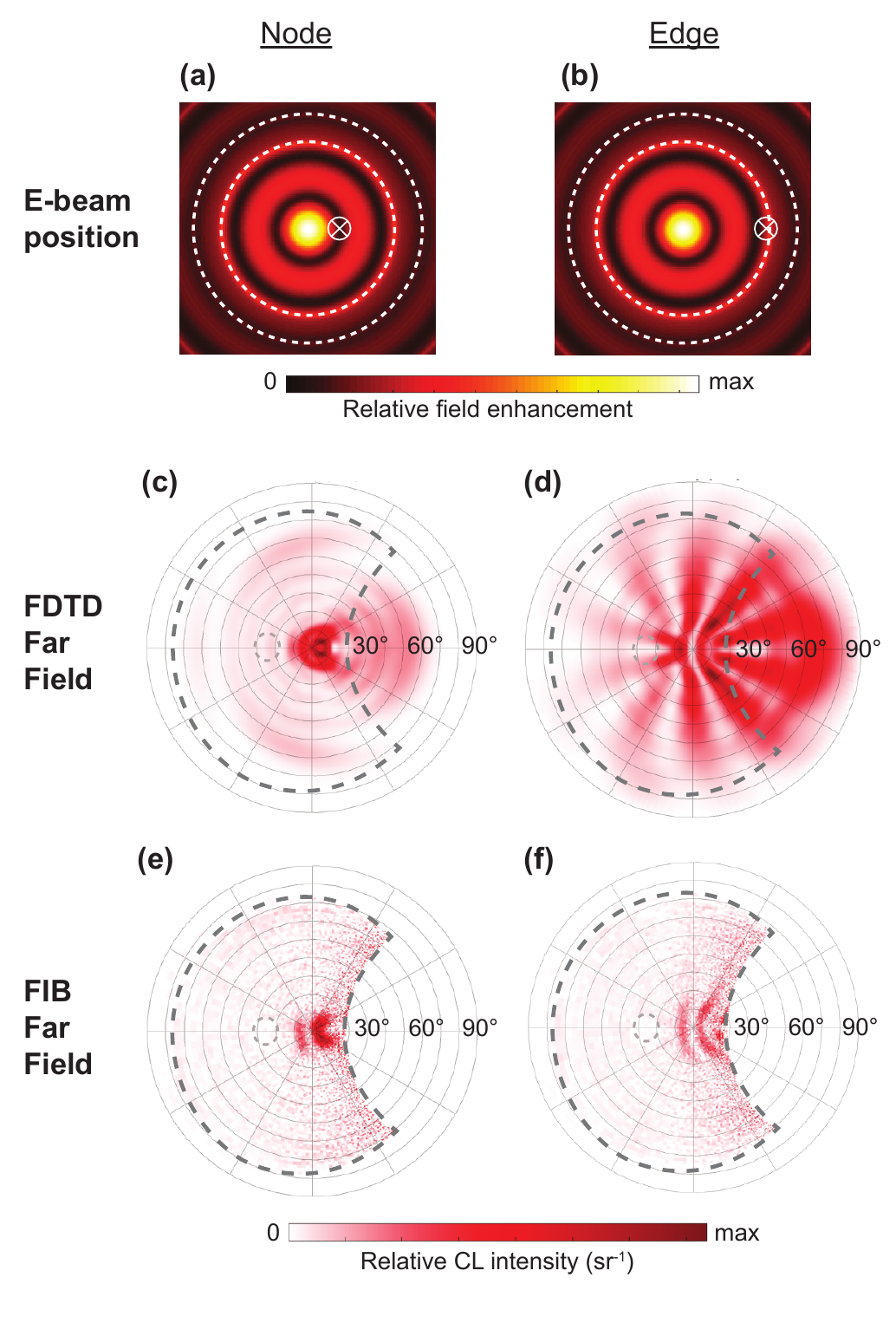}
\caption{Angle-resolved cathodoluminescence (CL) for a 4-ring lens when the electron beam is positioned off center. \textbf{a-b} Electron beam position (cross symbol) superimposed on the finite-difference time-domain (FDTD) simulated field enhancement profile for a radially polarized beam with 800 nm wavelength. The dashed circles indicate the edges of the first groove. \textbf{c-d} FDTD simulated far field polar maps for the electron beam at the node and edge positions. The gray dashed lines trace the electron beam entry hole and the open face of the parabolic mirror. \textbf{e-f} Measured CL far field polar maps at these positions for a lens fabricated using focused ion beam (FIB).}
\label{fig:CLFarFieldOffCenter}
\end{figure}

As discussed in section IV, scanning CL can be used to obtain qualitative maps of the plasmonic resonance and, more generally, the radiative LDOS. These are fundamentally different from the field enhancement profile expected upon excitation with a radially polarized beam. The main difference is that the electron beam only excites purely radial plasmon modes in the structure when the beam is positioned at the center. Elsewhere, the beam generates SPPs traveling outward from its current position rather than the lens center, so the modes excited do not match the circular symmetry of the lens. This plasmonic configuration would not be excited by a radially polarized laser. 

Using angle-resolved CL, we can observe this symmetry mismatch between plasmons and lens for off-center electron beam positions, shown in Figure~\ref{fig:CLFarFieldOffCenter} for a 4-ring lens. Instead of a circularly symmetric donut-shaped emission, complex interference patterns are generated in the far field whose shape depends on the beam position~\cite{polmanBullseyeCLPolarimetry}. 

For this reason, it is hard to quantitatively compare spatial profiles from CL spectromicroscopy and simulated field enhancement under laser illumination. One example of this is the observation that the CL emission is brighter when the beam is at the groove edge than at the center of the structure (see Figure~\ref{fig:ScanningCL}). At the edge, the plasmon modes excited and how they couple to the structure are much different than when the beam is at the center, as evidenced by the far field shown in Figure~\ref{fig:CLFarFieldOffCenter}d,f. Also, the edge is a strong scattering site, allowing many SPP modes generated nearby to couple to light, giving an enhanced radiative LDOS. Therefore, we still expect that the field enhancement and consequent photoemission under laser illumination will be stronger at the center than the edges. 

Quantitative analysis of the spatial map is further complicated by the nature of the CL process. The total measured intensity of this emission depends on how the emission interferes in the far field and what parts of the emission are collected by the mirror. At both the node and edge positions, some of the emitted light is not collected by the mirror (see Figure~\ref{fig:CLFarFieldOffCenter}c-f). Still, valuable qualitative insight can be obtained from scanning CL as discussed in Section IVB, such as visualizing the plasmonic resonance mode and nanoscale heterogeneities.

\bibliography{main}

\begin{thebibliography}{44}%
\makeatletter
\providecommand \@ifxundefined [1]{%
 \@ifx{#1\undefined}
}%
\providecommand \@ifnum [1]{%
 \ifnum #1\expandafter \@firstoftwo
 \else \expandafter \@secondoftwo
 \fi
}%
\providecommand \@ifx [1]{%
 \ifx #1\expandafter \@firstoftwo
 \else \expandafter \@secondoftwo
 \fi
}%
\providecommand \natexlab [1]{#1}%
\providecommand \enquote  [1]{``#1''}%
\providecommand \bibnamefont  [1]{#1}%
\providecommand \bibfnamefont [1]{#1}%
\providecommand \citenamefont [1]{#1}%
\providecommand \href@noop [0]{\@secondoftwo}%
\providecommand \href [0]{\begingroup \@sanitize@url \@href}%
\providecommand \@href[1]{\@@startlink{#1}\@@href}%
\providecommand \@@href[1]{\endgroup#1\@@endlink}%
\providecommand \@sanitize@url [0]{\catcode `\\12\catcode `\$12\catcode
  `\&12\catcode `\#12\catcode `\^12\catcode `\_12\catcode `\%12\relax}%
\providecommand \@@startlink[1]{}%
\providecommand \@@endlink[0]{}%
\providecommand \url  [0]{\begingroup\@sanitize@url \@url }%
\providecommand \@url [1]{\endgroup\@href {#1}{\urlprefix }}%
\providecommand \urlprefix  [0]{URL }%
\providecommand \Eprint [0]{\href }%
\providecommand \doibase [0]{http://dx.doi.org/}%
\providecommand \selectlanguage [0]{\@gobble}%
\providecommand \bibinfo  [0]{\@secondoftwo}%
\providecommand \bibfield  [0]{\@secondoftwo}%
\providecommand \translation [1]{[#1]}%
\providecommand \BibitemOpen [0]{}%
\providecommand \bibitemStop [0]{}%
\providecommand \bibitemNoStop [0]{.\EOS\space}%
\providecommand \EOS [0]{\spacefactor3000\relax}%
\providecommand \BibitemShut  [1]{\csname bibitem#1\endcsname}%
\let\auto@bib@innerbib\@empty
\bibitem [{\citenamefont {Siwick}\ \emph {et~al.}(2003)\citenamefont {Siwick},
  \citenamefont {Dwyer}, \citenamefont {Jordan},\ and\ \citenamefont
  {Miller}}]{siwick2003AlMelting}%
  \BibitemOpen
  \bibfield  {author} {\bibinfo {author} {\bibfnamefont {Bradley~J}\
  \bibnamefont {Siwick}}, \bibinfo {author} {\bibfnamefont {Jason~R}\
  \bibnamefont {Dwyer}}, \bibinfo {author} {\bibfnamefont {Robert~E}\
  \bibnamefont {Jordan}}, \ and\ \bibinfo {author} {\bibfnamefont {RJ~Dwayne}\
  \bibnamefont {Miller}},\ }\bibfield  {title} {\enquote {\bibinfo {title} {An
  atomic-level view of melting using femtosecond electron diffraction},}\
  }\href@noop {} {\bibfield  {journal} {\bibinfo  {journal} {Science}\ }\textbf
  {\bibinfo {volume} {302}},\ \bibinfo {pages} {1382--1385} (\bibinfo {year}
  {2003})}\BibitemShut {NoStop}%
\bibitem [{\citenamefont {Morrison}\ \emph {et~al.}(2014)\citenamefont
  {Morrison}, \citenamefont {Chatelain}, \citenamefont {Tiwari}, \citenamefont
  {Hendaoui}, \citenamefont {Bruh{\'a}cs}, \citenamefont {Chaker},\ and\
  \citenamefont {Siwick}}]{siwick2014VO2}%
  \BibitemOpen
  \bibfield  {author} {\bibinfo {author} {\bibfnamefont {Vance~R}\ \bibnamefont
  {Morrison}}, \bibinfo {author} {\bibfnamefont {Robert~P}\ \bibnamefont
  {Chatelain}}, \bibinfo {author} {\bibfnamefont {Kunal~L}\ \bibnamefont
  {Tiwari}}, \bibinfo {author} {\bibfnamefont {Ali}\ \bibnamefont {Hendaoui}},
  \bibinfo {author} {\bibfnamefont {Andrew}\ \bibnamefont {Bruh{\'a}cs}},
  \bibinfo {author} {\bibfnamefont {Mohamed}\ \bibnamefont {Chaker}}, \ and\
  \bibinfo {author} {\bibfnamefont {Bradley~J}\ \bibnamefont {Siwick}},\
  }\bibfield  {title} {\enquote {\bibinfo {title} {A photoinduced metal-like
  phase of monoclinic {VO$_{2}$} revealed by ultrafast electron diffraction},}\
  }\href@noop {} {\bibfield  {journal} {\bibinfo  {journal} {Science}\ }\textbf
  {\bibinfo {volume} {346}},\ \bibinfo {pages} {445--448} (\bibinfo {year}
  {2014})}\BibitemShut {NoStop}%
\bibitem [{\citenamefont {Eichberger}\ \emph {et~al.}(2010)\citenamefont
  {Eichberger}, \citenamefont {Sch{\"a}fer}, \citenamefont {Krumova},
  \citenamefont {Beyer}, \citenamefont {Demsar}, \citenamefont {Berger},
  \citenamefont {Moriena}, \citenamefont {Sciaini},\ and\ \citenamefont
  {Miller}}]{dwaynemiller2010TaS2}%
  \BibitemOpen
  \bibfield  {author} {\bibinfo {author} {\bibfnamefont {Maximilian}\
  \bibnamefont {Eichberger}}, \bibinfo {author} {\bibfnamefont {Hanjo}\
  \bibnamefont {Sch{\"a}fer}}, \bibinfo {author} {\bibfnamefont {Marina}\
  \bibnamefont {Krumova}}, \bibinfo {author} {\bibfnamefont {Markus}\
  \bibnamefont {Beyer}}, \bibinfo {author} {\bibfnamefont {Jure}\ \bibnamefont
  {Demsar}}, \bibinfo {author} {\bibfnamefont {Helmuth}\ \bibnamefont
  {Berger}}, \bibinfo {author} {\bibfnamefont {Gustavo}\ \bibnamefont
  {Moriena}}, \bibinfo {author} {\bibfnamefont {Germ{\'a}n}\ \bibnamefont
  {Sciaini}}, \ and\ \bibinfo {author} {\bibfnamefont {RJ~Dwayne}\ \bibnamefont
  {Miller}},\ }\bibfield  {title} {\enquote {\bibinfo {title} {Snapshots of
  cooperative atomic motions in the optical suppression of charge density
  waves},}\ }\href@noop {} {\bibfield  {journal} {\bibinfo  {journal} {Nature}\
  }\textbf {\bibinfo {volume} {468}},\ \bibinfo {pages} {799} (\bibinfo {year}
  {2010})}\BibitemShut {NoStop}%
\bibitem [{\citenamefont {Zewail}(2006)}]{zewail2006review}%
  \BibitemOpen
  \bibfield  {author} {\bibinfo {author} {\bibfnamefont {Ahmed~H}\ \bibnamefont
  {Zewail}},\ }\bibfield  {title} {\enquote {\bibinfo {title} {{4D} ultrafast
  electron diffraction, crystallography, and microscopy},}\ }\href@noop {}
  {\bibfield  {journal} {\bibinfo  {journal} {Annu. Rev. Phys. Chem.}\ }\textbf
  {\bibinfo {volume} {57}},\ \bibinfo {pages} {65--103} (\bibinfo {year}
  {2006})}\BibitemShut {NoStop}%
\bibitem [{\citenamefont {Gao}\ \emph {et~al.}(2013)\citenamefont {Gao},
  \citenamefont {Lu}, \citenamefont {Jean-Ruel}, \citenamefont {Liu},
  \citenamefont {Marx}, \citenamefont {Onda}, \citenamefont {Koshihara},
  \citenamefont {Nakano}, \citenamefont {Shao}, \citenamefont {Hiramatsu} \emph
  {et~al.}}]{dwaynemiller2013MITOrganicSalt}%
  \BibitemOpen
  \bibfield  {author} {\bibinfo {author} {\bibfnamefont {Meng}\ \bibnamefont
  {Gao}}, \bibinfo {author} {\bibfnamefont {Cheng}\ \bibnamefont {Lu}},
  \bibinfo {author} {\bibfnamefont {Hubert}\ \bibnamefont {Jean-Ruel}},
  \bibinfo {author} {\bibfnamefont {Lai~Chung}\ \bibnamefont {Liu}}, \bibinfo
  {author} {\bibfnamefont {Alexander}\ \bibnamefont {Marx}}, \bibinfo {author}
  {\bibfnamefont {Ken}\ \bibnamefont {Onda}}, \bibinfo {author} {\bibfnamefont
  {Shin-ya}\ \bibnamefont {Koshihara}}, \bibinfo {author} {\bibfnamefont
  {Yoshiaki}\ \bibnamefont {Nakano}}, \bibinfo {author} {\bibfnamefont
  {Xiangfeng}\ \bibnamefont {Shao}}, \bibinfo {author} {\bibfnamefont
  {Takaaki}\ \bibnamefont {Hiramatsu}},  \emph {et~al.},\ }\bibfield  {title}
  {\enquote {\bibinfo {title} {Mapping molecular motions leading to charge
  delocalization with ultrabright electrons},}\ }\href@noop {} {\bibfield
  {journal} {\bibinfo  {journal} {Nature}\ }\textbf {\bibinfo {volume} {496}},\
  \bibinfo {pages} {343} (\bibinfo {year} {2013})}\BibitemShut {NoStop}%
\bibitem [{\citenamefont {Rhee}(1992)}]{brightness}%
  \BibitemOpen
  \bibfield  {author} {\bibinfo {author} {\bibfnamefont {M.~J.}\ \bibnamefont
  {Rhee}},\ }\bibfield  {title} {\enquote {\bibinfo {title} {Refined definition
  of the beam brightness},}\ }\href {\doibase 10.1063/1.860076} {\bibfield
  {journal} {\bibinfo  {journal} {Physics of Fluids B: Plasma Physics}\
  }\textbf {\bibinfo {volume} {4}},\ \bibinfo {pages} {1674--1676} (\bibinfo
  {year} {1992})}\BibitemShut {NoStop}%
\bibitem [{\citenamefont {Carbone}\ \emph {et~al.}(2012)\citenamefont
  {Carbone}, \citenamefont {Musumeci}, \citenamefont {Luiten},\ and\
  \citenamefont {Hebert}}]{UEDperspective_2012}%
  \BibitemOpen
  \bibfield  {author} {\bibinfo {author} {\bibfnamefont {F.}~\bibnamefont
  {Carbone}}, \bibinfo {author} {\bibfnamefont {P.}~\bibnamefont {Musumeci}},
  \bibinfo {author} {\bibfnamefont {O.J.}\ \bibnamefont {Luiten}}, \ and\
  \bibinfo {author} {\bibfnamefont {C.}~\bibnamefont {Hebert}},\ }\bibfield
  {title} {\enquote {\bibinfo {title} {A perspective on novel sources of
  ultrashort electron and {X}-ray pulses},}\ }\href {\doibase
  10.1016/j.chemphys.2011.10.010} {\bibfield  {journal} {\bibinfo  {journal}
  {Chemical Physics}\ }\textbf {\bibinfo {volume} {392}},\ \bibinfo {pages}
  {1--9} (\bibinfo {year} {2012})}\BibitemShut {NoStop}%
\bibitem [{\citenamefont {Feist}\ \emph {et~al.}(2018)\citenamefont {Feist},
  \citenamefont {Rubiano~da Silva}, \citenamefont {Liang}, \citenamefont
  {Ropers},\ and\ \citenamefont {Sch{\"a}fer}}]{ropers2018UCBED}%
  \BibitemOpen
  \bibfield  {author} {\bibinfo {author} {\bibfnamefont {Armin}\ \bibnamefont
  {Feist}}, \bibinfo {author} {\bibfnamefont {Nara}\ \bibnamefont {Rubiano~da
  Silva}}, \bibinfo {author} {\bibfnamefont {Wenxi}\ \bibnamefont {Liang}},
  \bibinfo {author} {\bibfnamefont {Claus}\ \bibnamefont {Ropers}}, \ and\
  \bibinfo {author} {\bibfnamefont {Sascha}\ \bibnamefont {Sch{\"a}fer}},\
  }\bibfield  {title} {\enquote {\bibinfo {title} {Nanoscale diffractive
  probing of strain dynamics in ultrafast transmission electron microscopy},}\
  }\href@noop {} {\bibfield  {journal} {\bibinfo  {journal} {Structural
  Dynamics}\ }\textbf {\bibinfo {volume} {5}},\ \bibinfo {pages} {014302}
  (\bibinfo {year} {2018})}\BibitemShut {NoStop}%
\bibitem [{\citenamefont {da~Silva}\ \emph {et~al.}(2018)\citenamefont
  {da~Silva}, \citenamefont {M{\"o}ller}, \citenamefont {Feist}, \citenamefont
  {Ulrichs}, \citenamefont {Ropers},\ and\ \citenamefont
  {Sch{\"a}fer}}]{ropers2018ULTEM}%
  \BibitemOpen
  \bibfield  {author} {\bibinfo {author} {\bibfnamefont {Nara~Rubiano}\
  \bibnamefont {da~Silva}}, \bibinfo {author} {\bibfnamefont {Marcel}\
  \bibnamefont {M{\"o}ller}}, \bibinfo {author} {\bibfnamefont {Armin}\
  \bibnamefont {Feist}}, \bibinfo {author} {\bibfnamefont {Henning}\
  \bibnamefont {Ulrichs}}, \bibinfo {author} {\bibfnamefont {Claus}\
  \bibnamefont {Ropers}}, \ and\ \bibinfo {author} {\bibfnamefont {Sascha}\
  \bibnamefont {Sch{\"a}fer}},\ }\bibfield  {title} {\enquote {\bibinfo {title}
  {Nanoscale mapping of ultrafast magnetization dynamics with femtosecond
  lorentz microscopy},}\ }\href@noop {} {\bibfield  {journal} {\bibinfo
  {journal} {Physical Review X}\ }\textbf {\bibinfo {volume} {8}},\ \bibinfo
  {pages} {031052} (\bibinfo {year} {2018})}\BibitemShut {NoStop}%
\bibitem [{\citenamefont {Yurtsever}\ \emph {et~al.}(2012)\citenamefont
  {Yurtsever}, \citenamefont {van~der Veen},\ and\ \citenamefont
  {Zewail}}]{zewail2012UEELSPlasmonic}%
  \BibitemOpen
  \bibfield  {author} {\bibinfo {author} {\bibfnamefont {Aycan}\ \bibnamefont
  {Yurtsever}}, \bibinfo {author} {\bibfnamefont {Renske~M}\ \bibnamefont
  {van~der Veen}}, \ and\ \bibinfo {author} {\bibfnamefont {Ahmed~H}\
  \bibnamefont {Zewail}},\ }\bibfield  {title} {\enquote {\bibinfo {title}
  {Subparticle ultrafast spectrum imaging in {4D} electron microscopy},}\
  }\href@noop {} {\bibfield  {journal} {\bibinfo  {journal} {Science}\ }\textbf
  {\bibinfo {volume} {335}},\ \bibinfo {pages} {59--64} (\bibinfo {year}
  {2012})}\BibitemShut {NoStop}%
\bibitem [{\citenamefont {Maxson}\ \emph {et~al.}(2017)\citenamefont {Maxson},
  \citenamefont {Cesar}, \citenamefont {Calmasini}, \citenamefont {Ody},
  \citenamefont {Musumeci},\ and\ \citenamefont
  {Alesini}}]{maxson_direct_2017}%
  \BibitemOpen
  \bibfield  {author} {\bibinfo {author} {\bibfnamefont {Jared}\ \bibnamefont
  {Maxson}}, \bibinfo {author} {\bibfnamefont {David}\ \bibnamefont {Cesar}},
  \bibinfo {author} {\bibfnamefont {Giacomo}\ \bibnamefont {Calmasini}},
  \bibinfo {author} {\bibfnamefont {Alexander}\ \bibnamefont {Ody}}, \bibinfo
  {author} {\bibfnamefont {Pietro}\ \bibnamefont {Musumeci}}, \ and\ \bibinfo
  {author} {\bibfnamefont {David}\ \bibnamefont {Alesini}},\ }\bibfield
  {title} {\enquote {\bibinfo {title} {Direct measurement of sub-10 fs
  relativistic electron beams with ultralow emittance},}\ }\href {\doibase
  10.1103/PhysRevLett.118.154802} {\bibfield  {journal} {\bibinfo  {journal}
  {Physical Review Letters}\ }\textbf {\bibinfo {volume} {118}} (\bibinfo
  {year} {2017}),\ 10.1103/PhysRevLett.118.154802}\BibitemShut {NoStop}%
\bibitem [{\citenamefont {Zhao}\ \emph {et~al.}(2018)\citenamefont {Zhao},
  \citenamefont {Wang}, \citenamefont {Lu}, \citenamefont {Wang}, \citenamefont
  {Hu}, \citenamefont {Wang}, \citenamefont {Qi}, \citenamefont {Jiang},
  \citenamefont {Liu}, \citenamefont {Ma}, \citenamefont {Qi}, \citenamefont
  {Zhu}, \citenamefont {Cheng}, \citenamefont {Shi}, \citenamefont {Shi},
  \citenamefont {Song}, \citenamefont {Zhu}, \citenamefont {Shi}, \citenamefont
  {Wang}, \citenamefont {Yan}, \citenamefont {Zhu}, \citenamefont {Xiang},\
  and\ \citenamefont {Zhang}}]{zhao_terahertz_2018}%
  \BibitemOpen
  \bibfield  {author} {\bibinfo {author} {\bibfnamefont {Lingrong}\
  \bibnamefont {Zhao}}, \bibinfo {author} {\bibfnamefont {Zhe}\ \bibnamefont
  {Wang}}, \bibinfo {author} {\bibfnamefont {Chao}\ \bibnamefont {Lu}},
  \bibinfo {author} {\bibfnamefont {Rui}\ \bibnamefont {Wang}}, \bibinfo
  {author} {\bibfnamefont {Cheng}\ \bibnamefont {Hu}}, \bibinfo {author}
  {\bibfnamefont {Peng}\ \bibnamefont {Wang}}, \bibinfo {author} {\bibfnamefont
  {Jia}\ \bibnamefont {Qi}}, \bibinfo {author} {\bibfnamefont {Tao}\
  \bibnamefont {Jiang}}, \bibinfo {author} {\bibfnamefont {Shengguang}\
  \bibnamefont {Liu}}, \bibinfo {author} {\bibfnamefont {Zhuoran}\ \bibnamefont
  {Ma}}, \bibinfo {author} {\bibfnamefont {Fengfeng}\ \bibnamefont {Qi}},
  \bibinfo {author} {\bibfnamefont {Pengfei}\ \bibnamefont {Zhu}}, \bibinfo
  {author} {\bibfnamefont {Ya}~\bibnamefont {Cheng}}, \bibinfo {author}
  {\bibfnamefont {Zhiwen}\ \bibnamefont {Shi}}, \bibinfo {author}
  {\bibfnamefont {Yanchao}\ \bibnamefont {Shi}}, \bibinfo {author}
  {\bibfnamefont {Wei}\ \bibnamefont {Song}}, \bibinfo {author} {\bibfnamefont
  {Xiaoxin}\ \bibnamefont {Zhu}}, \bibinfo {author} {\bibfnamefont {Jiaru}\
  \bibnamefont {Shi}}, \bibinfo {author} {\bibfnamefont {Yingxin}\ \bibnamefont
  {Wang}}, \bibinfo {author} {\bibfnamefont {Lixin}\ \bibnamefont {Yan}},
  \bibinfo {author} {\bibfnamefont {Liguo}\ \bibnamefont {Zhu}}, \bibinfo
  {author} {\bibfnamefont {Dao}\ \bibnamefont {Xiang}}, \ and\ \bibinfo
  {author} {\bibfnamefont {Jie}\ \bibnamefont {Zhang}},\ }\bibfield  {title}
  {\enquote {\bibinfo {title} {Terahertz streaking of few-femtosecond
  relativistic electron beams},}\ }\href {\doibase 10.1103/PhysRevX.8.021061}
  {\bibfield  {journal} {\bibinfo  {journal} {Physical Review X}\ }\textbf
  {\bibinfo {volume} {8}} (\bibinfo {year} {2018}),\
  10.1103/PhysRevX.8.021061}\BibitemShut {NoStop}%
\bibitem [{\citenamefont {Huang}\ \emph {et~al.}(2015)\citenamefont {Huang},
  \citenamefont {Filippetto}, \citenamefont {Papadopoulos}, \citenamefont
  {Qian}, \citenamefont {Sannibale},\ and\ \citenamefont
  {Zolotorev}}]{DarkCurrentApex}%
  \BibitemOpen
  \bibfield  {author} {\bibinfo {author} {\bibfnamefont {R.}~\bibnamefont
  {Huang}}, \bibinfo {author} {\bibfnamefont {D.}~\bibnamefont {Filippetto}},
  \bibinfo {author} {\bibfnamefont {C.F.}\ \bibnamefont {Papadopoulos}},
  \bibinfo {author} {\bibfnamefont {H.}~\bibnamefont {Qian}}, \bibinfo {author}
  {\bibfnamefont {F.}~\bibnamefont {Sannibale}}, \ and\ \bibinfo {author}
  {\bibfnamefont {M.}~\bibnamefont {Zolotorev}},\ }\bibfield  {title} {\enquote
  {\bibinfo {title} {Dark current studies on a normal-conducting
  high-brightness very-high-frequency electron gun operating in continuous wave
  mode},}\ }\href {\doibase 10.1103/PhysRevSTAB.18.013401} {\bibfield
  {journal} {\bibinfo  {journal} {Physical Review Special Topics - Accelerators
  and Beams}\ }\textbf {\bibinfo {volume} {18}} (\bibinfo {year} {2015}),\
  10.1103/PhysRevSTAB.18.013401}\BibitemShut {NoStop}%
\bibitem [{\citenamefont {Polyakov}\ \emph {et~al.}(2013)\citenamefont
  {Polyakov}, \citenamefont {Senft}, \citenamefont {Thompson}, \citenamefont
  {Feng}, \citenamefont {Cabrini}, \citenamefont {Schuck}, \citenamefont
  {Padmore}, \citenamefont {Peppernick},\ and\ \citenamefont
  {Hess}}]{polyakov2013nanogrooveCathodes}%
  \BibitemOpen
  \bibfield  {author} {\bibinfo {author} {\bibfnamefont {Aleksandr}\
  \bibnamefont {Polyakov}}, \bibinfo {author} {\bibfnamefont {Christoph}\
  \bibnamefont {Senft}}, \bibinfo {author} {\bibfnamefont {KF}~\bibnamefont
  {Thompson}}, \bibinfo {author} {\bibfnamefont {J}~\bibnamefont {Feng}},
  \bibinfo {author} {\bibfnamefont {S}~\bibnamefont {Cabrini}}, \bibinfo
  {author} {\bibfnamefont {PJ}~\bibnamefont {Schuck}}, \bibinfo {author}
  {\bibfnamefont {HA}~\bibnamefont {Padmore}}, \bibinfo {author} {\bibfnamefont
  {Samuel~J}\ \bibnamefont {Peppernick}}, \ and\ \bibinfo {author}
  {\bibfnamefont {Wayne~P}\ \bibnamefont {Hess}},\ }\bibfield  {title}
  {\enquote {\bibinfo {title} {Plasmon-enhanced photocathode for high
  brightness and high repetition rate x-ray sources},}\ }\href@noop {}
  {\bibfield  {journal} {\bibinfo  {journal} {Physical review letters}\
  }\textbf {\bibinfo {volume} {110}},\ \bibinfo {pages} {076802} (\bibinfo
  {year} {2013})}\BibitemShut {NoStop}%
\bibitem [{\citenamefont {Li}\ \emph {et~al.}(2013)\citenamefont {Li},
  \citenamefont {To}, \citenamefont {Andonian}, \citenamefont {Feng},
  \citenamefont {Polyakov}, \citenamefont {Scoby}, \citenamefont {Thompson},
  \citenamefont {Wan}, \citenamefont {Padmore},\ and\ \citenamefont
  {Musumeci}}]{Musumeci2013nanoholeCathode}%
  \BibitemOpen
  \bibfield  {author} {\bibinfo {author} {\bibfnamefont {RK}~\bibnamefont
  {Li}}, \bibinfo {author} {\bibfnamefont {H}~\bibnamefont {To}}, \bibinfo
  {author} {\bibfnamefont {G}~\bibnamefont {Andonian}}, \bibinfo {author}
  {\bibfnamefont {J}~\bibnamefont {Feng}}, \bibinfo {author} {\bibfnamefont
  {A}~\bibnamefont {Polyakov}}, \bibinfo {author} {\bibfnamefont
  {CM}~\bibnamefont {Scoby}}, \bibinfo {author} {\bibfnamefont {K}~\bibnamefont
  {Thompson}}, \bibinfo {author} {\bibfnamefont {W}~\bibnamefont {Wan}},
  \bibinfo {author} {\bibfnamefont {HA}~\bibnamefont {Padmore}}, \ and\
  \bibinfo {author} {\bibfnamefont {P}~\bibnamefont {Musumeci}},\ }\bibfield
  {title} {\enquote {\bibinfo {title} {Surface-plasmon resonance-enhanced
  multiphoton emission of high-brightness electron beams from a nanostructured
  copper cathode},}\ }\href@noop {} {\bibfield  {journal} {\bibinfo  {journal}
  {Physical review letters}\ }\textbf {\bibinfo {volume} {110}},\ \bibinfo
  {pages} {074801} (\bibinfo {year} {2013})}\BibitemShut {NoStop}%
\bibitem [{\citenamefont {Lau}(1987)}]{roughness}%
  \BibitemOpen
  \bibfield  {author} {\bibinfo {author} {\bibfnamefont {Y.~Y.}\ \bibnamefont
  {Lau}},\ }\bibfield  {title} {\enquote {\bibinfo {title} {Effects of cathode
  surface roughness on the quality of electron beams},}\ }\href {\doibase
  10.1063/1.338833} {\bibfield  {journal} {\bibinfo  {journal} {Journal of
  Applied Physics}\ }\textbf {\bibinfo {volume} {61}},\ \bibinfo {pages}
  {36--44} (\bibinfo {year} {1987})},\ \Eprint
  {http://arxiv.org/abs/https://doi.org/10.1063/1.338833}
  {https://doi.org/10.1063/1.338833} \BibitemShut {NoStop}%
\bibitem [{\citenamefont {Steele}\ \emph {et~al.}(2006)\citenamefont {Steele},
  \citenamefont {Liu}, \citenamefont {Wang},\ and\ \citenamefont
  {Zhang}}]{zhang2006linPolBESNOM}%
  \BibitemOpen
  \bibfield  {author} {\bibinfo {author} {\bibfnamefont {Jennifer~M}\
  \bibnamefont {Steele}}, \bibinfo {author} {\bibfnamefont {Zhaowei}\
  \bibnamefont {Liu}}, \bibinfo {author} {\bibfnamefont {Yuan}\ \bibnamefont
  {Wang}}, \ and\ \bibinfo {author} {\bibfnamefont {Xiang}\ \bibnamefont
  {Zhang}},\ }\bibfield  {title} {\enquote {\bibinfo {title} {Resonant and
  non-resonant generation and focusing of surface plasmons with circular
  gratings},}\ }\href@noop {} {\bibfield  {journal} {\bibinfo  {journal}
  {Optics Express}\ }\textbf {\bibinfo {volume} {14}},\ \bibinfo {pages}
  {5664--5670} (\bibinfo {year} {2006})}\BibitemShut {NoStop}%
\bibitem [{\citenamefont {Chen}\ \emph {et~al.}(2009)\citenamefont {Chen},
  \citenamefont {Abeysinghe}, \citenamefont {Nelson},\ and\ \citenamefont
  {Zhan}}]{chen2009radialPolarizationBE}%
  \BibitemOpen
  \bibfield  {author} {\bibinfo {author} {\bibfnamefont {Weibin}\ \bibnamefont
  {Chen}}, \bibinfo {author} {\bibfnamefont {Don~C}\ \bibnamefont
  {Abeysinghe}}, \bibinfo {author} {\bibfnamefont {Robert~L}\ \bibnamefont
  {Nelson}}, \ and\ \bibinfo {author} {\bibfnamefont {Qiwen}\ \bibnamefont
  {Zhan}},\ }\bibfield  {title} {\enquote {\bibinfo {title} {Plasmonic lens
  made of multiple concentric metallic rings under radially polarized
  illumination},}\ }\href@noop {} {\bibfield  {journal} {\bibinfo  {journal}
  {Nano letters}\ }\textbf {\bibinfo {volume} {9}},\ \bibinfo {pages}
  {4320--4325} (\bibinfo {year} {2009})}\BibitemShut {NoStop}%
\bibitem [{\citenamefont {Bechtel}\ \emph {et~al.}(1977)\citenamefont
  {Bechtel}, \citenamefont {Smith},\ and\ \citenamefont
  {Bloembergen}}]{bechtel1977generalizedFD}%
  \BibitemOpen
  \bibfield  {author} {\bibinfo {author} {\bibfnamefont {JH}~\bibnamefont
  {Bechtel}}, \bibinfo {author} {\bibfnamefont {W~Lee}\ \bibnamefont {Smith}},
  \ and\ \bibinfo {author} {\bibfnamefont {N}~\bibnamefont {Bloembergen}},\
  }\bibfield  {title} {\enquote {\bibinfo {title} {Two-photon photoemission
  from metals induced by picosecond laser pulses},}\ }\href@noop {} {\bibfield
  {journal} {\bibinfo  {journal} {Physical Review B}\ }\textbf {\bibinfo
  {volume} {15}},\ \bibinfo {pages} {4557} (\bibinfo {year}
  {1977})}\BibitemShut {NoStop}%
\bibitem [{\citenamefont {Kolomenski}\ \emph {et~al.}(2009)\citenamefont
  {Kolomenski}, \citenamefont {Kolomenskii}, \citenamefont {Noel},
  \citenamefont {Peng},\ and\ \citenamefont
  {Schuessler}}]{schuessler2009AuSPPPropLength}%
  \BibitemOpen
  \bibfield  {author} {\bibinfo {author} {\bibfnamefont {Andrei}\ \bibnamefont
  {Kolomenski}}, \bibinfo {author} {\bibfnamefont {Alexandre}\ \bibnamefont
  {Kolomenskii}}, \bibinfo {author} {\bibfnamefont {John}\ \bibnamefont
  {Noel}}, \bibinfo {author} {\bibfnamefont {Siying}\ \bibnamefont {Peng}}, \
  and\ \bibinfo {author} {\bibfnamefont {Hans}\ \bibnamefont {Schuessler}},\
  }\bibfield  {title} {\enquote {\bibinfo {title} {Propagation length of
  surface plasmons in a metal film with roughness},}\ }\href@noop {} {\bibfield
   {journal} {\bibinfo  {journal} {Applied optics}\ }\textbf {\bibinfo {volume}
  {48}},\ \bibinfo {pages} {5683--5691} (\bibinfo {year} {2009})}\BibitemShut
  {NoStop}%
\bibitem [{\citenamefont {Kuttge}\ \emph {et~al.}(2008)\citenamefont {Kuttge},
  \citenamefont {Vesseur}, \citenamefont {Verhoeven}, \citenamefont {Lezec},
  \citenamefont {Atwater},\ and\ \citenamefont {Polman}}]{kuttge2008loss}%
  \BibitemOpen
  \bibfield  {author} {\bibinfo {author} {\bibfnamefont {M}~\bibnamefont
  {Kuttge}}, \bibinfo {author} {\bibfnamefont {EJR}\ \bibnamefont {Vesseur}},
  \bibinfo {author} {\bibfnamefont {J}~\bibnamefont {Verhoeven}}, \bibinfo
  {author} {\bibfnamefont {HJ}~\bibnamefont {Lezec}}, \bibinfo {author}
  {\bibfnamefont {HA}~\bibnamefont {Atwater}}, \ and\ \bibinfo {author}
  {\bibfnamefont {A}~\bibnamefont {Polman}},\ }\bibfield  {title} {\enquote
  {\bibinfo {title} {Loss mechanisms of surface plasmon polaritons on gold
  probed by cathodoluminescence imaging spectroscopy.}}\ }\href@noop {}
  {\bibfield  {journal} {\bibinfo  {journal} {Applied Physics Letters}\
  }\textbf {\bibinfo {volume} {93}},\ \bibinfo {pages} {113110} (\bibinfo
  {year} {2008})}\BibitemShut {NoStop}%
\bibitem [{\citenamefont {McPeak}\ \emph {et~al.}(2015)\citenamefont {McPeak},
  \citenamefont {Jayanti}, \citenamefont {Kress}, \citenamefont {Meyer},
  \citenamefont {Iotti}, \citenamefont {Rossinelli},\ and\ \citenamefont
  {Norris}}]{mcpeak2015plasmonic}%
  \BibitemOpen
  \bibfield  {author} {\bibinfo {author} {\bibfnamefont {Kevin~M}\ \bibnamefont
  {McPeak}}, \bibinfo {author} {\bibfnamefont {Sriharsha~V}\ \bibnamefont
  {Jayanti}}, \bibinfo {author} {\bibfnamefont {Stephan~JP}\ \bibnamefont
  {Kress}}, \bibinfo {author} {\bibfnamefont {Stefan}\ \bibnamefont {Meyer}},
  \bibinfo {author} {\bibfnamefont {Stelio}\ \bibnamefont {Iotti}}, \bibinfo
  {author} {\bibfnamefont {Aurelio}\ \bibnamefont {Rossinelli}}, \ and\
  \bibinfo {author} {\bibfnamefont {David~J}\ \bibnamefont {Norris}},\
  }\bibfield  {title} {\enquote {\bibinfo {title} {Plasmonic films can easily
  be better: rules and recipes},}\ }\href@noop {} {\bibfield  {journal}
  {\bibinfo  {journal} {ACS photonics}\ }\textbf {\bibinfo {volume} {2}},\
  \bibinfo {pages} {326--333} (\bibinfo {year} {2015})}\BibitemShut {NoStop}%
\bibitem [{\citenamefont {Olmon}\ \emph {et~al.}(2012)\citenamefont {Olmon},
  \citenamefont {Slovick}, \citenamefont {Johnson}, \citenamefont {Shelton},
  \citenamefont {Oh}, \citenamefont {Boreman},\ and\ \citenamefont
  {Raschke}}]{olmon2012optical}%
  \BibitemOpen
  \bibfield  {author} {\bibinfo {author} {\bibfnamefont {Robert~L}\
  \bibnamefont {Olmon}}, \bibinfo {author} {\bibfnamefont {Brian}\ \bibnamefont
  {Slovick}}, \bibinfo {author} {\bibfnamefont {Timothy~W}\ \bibnamefont
  {Johnson}}, \bibinfo {author} {\bibfnamefont {David}\ \bibnamefont
  {Shelton}}, \bibinfo {author} {\bibfnamefont {Sang-Hyun}\ \bibnamefont {Oh}},
  \bibinfo {author} {\bibfnamefont {Glenn~D}\ \bibnamefont {Boreman}}, \ and\
  \bibinfo {author} {\bibfnamefont {Markus~B}\ \bibnamefont {Raschke}},\
  }\bibfield  {title} {\enquote {\bibinfo {title} {Optical dielectric function
  of gold},}\ }\href@noop {} {\bibfield  {journal} {\bibinfo  {journal}
  {Physical Review B}\ }\textbf {\bibinfo {volume} {86}},\ \bibinfo {pages}
  {235147} (\bibinfo {year} {2012})}\BibitemShut {NoStop}%
\bibitem [{\citenamefont {Raether}(1988)}]{raetherBookSPsOnGratings}%
  \BibitemOpen
  \bibfield  {author} {\bibinfo {author} {\bibfnamefont {Heinz}\ \bibnamefont
  {Raether}},\ }\bibfield  {title} {\enquote {\bibinfo {title} {Surface
  plasmons on gratings},}\ }in\ \href@noop {} {\emph {\bibinfo {booktitle}
  {Surface plasmons on smooth and rough surfaces and on gratings}}}\ (\bibinfo
  {publisher} {Springer},\ \bibinfo {year} {1988})\ pp.\ \bibinfo {pages}
  {91--116}\BibitemShut {NoStop}%
\bibitem [{\citenamefont {Lumerical}()}]{Lumerical}%
  \BibitemOpen
  \bibfield  {author} {\bibinfo {author} {\bibnamefont {Lumerical}},\
  }\href@noop {} {\enquote {\bibinfo {title} {{FDTD} solutions},}\ }\bibinfo
  {howpublished} {\url{http://www.lumerical.com}}\BibitemShut {NoStop}%
\bibitem [{\citenamefont {Laughton}\ and\ \citenamefont
  {Warne}(2003)}]{warne2005handbook}%
  \BibitemOpen
  \bibfield  {author} {\bibinfo {author} {\bibfnamefont {M~A}\ \bibnamefont
  {Laughton}}\ and\ \bibinfo {author} {\bibfnamefont {D~F}\ \bibnamefont
  {Warne}},\ }\href@noop {} {\emph {\bibinfo {title} {Electrical engineer's
  reference book}}},\ \bibinfo {edition} {16th}\ ed.\ (\bibinfo  {publisher}
  {Elsevier},\ \bibinfo {year} {2003})\ pp.\ \bibinfo {pages}
  {384--387}\BibitemShut {NoStop}%
\bibitem [{\citenamefont {Thio}\ \emph {et~al.}(2001)\citenamefont {Thio},
  \citenamefont {Pellerin}, \citenamefont {Linke}, \citenamefont {Lezec},\ and\
  \citenamefont {Ebbesen}}]{ebbesen2001TransmissionBE}%
  \BibitemOpen
  \bibfield  {author} {\bibinfo {author} {\bibfnamefont {Tineke}\ \bibnamefont
  {Thio}}, \bibinfo {author} {\bibfnamefont {KM}~\bibnamefont {Pellerin}},
  \bibinfo {author} {\bibfnamefont {RA}~\bibnamefont {Linke}}, \bibinfo
  {author} {\bibfnamefont {HJ}~\bibnamefont {Lezec}}, \ and\ \bibinfo {author}
  {\bibfnamefont {TW}~\bibnamefont {Ebbesen}},\ }\bibfield  {title} {\enquote
  {\bibinfo {title} {Enhanced light transmission through a single subwavelength
  aperture},}\ }\href@noop {} {\bibfield  {journal} {\bibinfo  {journal}
  {Optics letters}\ }\textbf {\bibinfo {volume} {26}},\ \bibinfo {pages}
  {1972--1974} (\bibinfo {year} {2001})}\BibitemShut {NoStop}%
\bibitem [{\citenamefont {Hooper}\ and\ \citenamefont
  {Sambles}(2002)}]{hooper2002deepGratings}%
  \BibitemOpen
  \bibfield  {author} {\bibinfo {author} {\bibfnamefont {Ian~R}\ \bibnamefont
  {Hooper}}\ and\ \bibinfo {author} {\bibfnamefont {J~Roy}\ \bibnamefont
  {Sambles}},\ }\bibfield  {title} {\enquote {\bibinfo {title} {Dispersion of
  surface plasmon polaritons on short-pitch metal gratings},}\ }\href@noop {}
  {\bibfield  {journal} {\bibinfo  {journal} {Physical Review B}\ }\textbf
  {\bibinfo {volume} {65}},\ \bibinfo {pages} {165432} (\bibinfo {year}
  {2002})}\BibitemShut {NoStop}%
\bibitem [{\citenamefont {Vogel}\ \emph {et~al.}(2012)\citenamefont {Vogel},
  \citenamefont {Zieleniecki},\ and\ \citenamefont
  {K{\"o}per}}]{vogel2012flat}%
  \BibitemOpen
  \bibfield  {author} {\bibinfo {author} {\bibfnamefont {Nicolas}\ \bibnamefont
  {Vogel}}, \bibinfo {author} {\bibfnamefont {Julius}\ \bibnamefont
  {Zieleniecki}}, \ and\ \bibinfo {author} {\bibfnamefont {Ingo}\ \bibnamefont
  {K{\"o}per}},\ }\bibfield  {title} {\enquote {\bibinfo {title} {As flat as it
  gets: ultrasmooth surfaces from template-stripping procedures},}\ }\href@noop
  {} {\bibfield  {journal} {\bibinfo  {journal} {Nanoscale}\ }\textbf {\bibinfo
  {volume} {4}},\ \bibinfo {pages} {3820--3832} (\bibinfo {year}
  {2012})}\BibitemShut {NoStop}%
\bibitem [{\citenamefont {De~Abajo}(2010)}]{deAbajoOpticalExcitationsReview}%
  \BibitemOpen
  \bibfield  {author} {\bibinfo {author} {\bibfnamefont {FJ~Garc{\'\i}a}\
  \bibnamefont {De~Abajo}},\ }\bibfield  {title} {\enquote {\bibinfo {title}
  {Optical excitations in electron microscopy},}\ }\href@noop {} {\bibfield
  {journal} {\bibinfo  {journal} {Reviews of modern physics}\ }\textbf
  {\bibinfo {volume} {82}},\ \bibinfo {pages} {209} (\bibinfo {year}
  {2010})}\BibitemShut {NoStop}%
\bibitem [{\citenamefont {Kuttge}\ \emph {et~al.}(2009)\citenamefont {Kuttge},
  \citenamefont {Vesseur}, \citenamefont {Koenderink}, \citenamefont {Lezec},
  \citenamefont {Atwater}, \citenamefont {de~Abajo},\ and\ \citenamefont
  {Polman}}]{polmanCLLinearGrating}%
  \BibitemOpen
  \bibfield  {author} {\bibinfo {author} {\bibfnamefont {Martin}\ \bibnamefont
  {Kuttge}}, \bibinfo {author} {\bibfnamefont {Ernst Jan~R}\ \bibnamefont
  {Vesseur}}, \bibinfo {author} {\bibfnamefont {AF}~\bibnamefont {Koenderink}},
  \bibinfo {author} {\bibfnamefont {HJ}~\bibnamefont {Lezec}}, \bibinfo
  {author} {\bibfnamefont {HA}~\bibnamefont {Atwater}}, \bibinfo {author}
  {\bibfnamefont {FJ~Garc{\'\i}a}\ \bibnamefont {de~Abajo}}, \ and\ \bibinfo
  {author} {\bibfnamefont {Albert}\ \bibnamefont {Polman}},\ }\bibfield
  {title} {\enquote {\bibinfo {title} {Local density of states, spectrum, and
  far-field interference of surface plasmon polaritons probed by
  cathodoluminescence},}\ }\href@noop {} {\bibfield  {journal} {\bibinfo
  {journal} {Physical Review B}\ }\textbf {\bibinfo {volume} {79}},\ \bibinfo
  {pages} {113405} (\bibinfo {year} {2009})}\BibitemShut {NoStop}%
\bibitem [{\citenamefont {Hofmann}\ \emph {et~al.}(2007)\citenamefont
  {Hofmann}, \citenamefont {Vesseur}, \citenamefont {Sweatlock}, \citenamefont
  {Lezec}, \citenamefont {Garc{\'\i}a~de Abajo}, \citenamefont {Polman},\ and\
  \citenamefont {Atwater}}]{atwaterBullseyeCL}%
  \BibitemOpen
  \bibfield  {author} {\bibinfo {author} {\bibfnamefont {Carrie~E}\
  \bibnamefont {Hofmann}}, \bibinfo {author} {\bibfnamefont {Ernst Jan~R}\
  \bibnamefont {Vesseur}}, \bibinfo {author} {\bibfnamefont {Luke~A}\
  \bibnamefont {Sweatlock}}, \bibinfo {author} {\bibfnamefont {Henri~J}\
  \bibnamefont {Lezec}}, \bibinfo {author} {\bibfnamefont {F~Javier}\
  \bibnamefont {Garc{\'\i}a~de Abajo}}, \bibinfo {author} {\bibfnamefont
  {Albert}\ \bibnamefont {Polman}}, \ and\ \bibinfo {author} {\bibfnamefont
  {Harry~A}\ \bibnamefont {Atwater}},\ }\bibfield  {title} {\enquote {\bibinfo
  {title} {Plasmonic modes of annular nanoresonators imaged by spectrally
  resolved cathodoluminescence},}\ }\href@noop {} {\bibfield  {journal}
  {\bibinfo  {journal} {Nano letters}\ }\textbf {\bibinfo {volume} {7}},\
  \bibinfo {pages} {3612--3617} (\bibinfo {year} {2007})}\BibitemShut {NoStop}%
\bibitem [{\citenamefont {Losquin}\ and\ \citenamefont
  {Kociak}(2015)}]{kociak2015LDOS}%
  \BibitemOpen
  \bibfield  {author} {\bibinfo {author} {\bibfnamefont {Arthur}\ \bibnamefont
  {Losquin}}\ and\ \bibinfo {author} {\bibfnamefont {Mathieu}\ \bibnamefont
  {Kociak}},\ }\bibfield  {title} {\enquote {\bibinfo {title} {Link between
  cathodoluminescence and electron energy loss spectroscopy and the radiative
  and full electromagnetic local density of states},}\ }\href@noop {}
  {\bibfield  {journal} {\bibinfo  {journal} {ACS Photonics}\ }\textbf
  {\bibinfo {volume} {2}},\ \bibinfo {pages} {1619--1627} (\bibinfo {year}
  {2015})}\BibitemShut {NoStop}%
\bibitem [{\citenamefont {Coenen}\ \emph {et~al.}(2015)\citenamefont {Coenen},
  \citenamefont {Brenny}, \citenamefont {Vesseur},\ and\ \citenamefont
  {Polman}}]{polmanCLReview}%
  \BibitemOpen
  \bibfield  {author} {\bibinfo {author} {\bibfnamefont {Toon}\ \bibnamefont
  {Coenen}}, \bibinfo {author} {\bibfnamefont {Benjamin~JM}\ \bibnamefont
  {Brenny}}, \bibinfo {author} {\bibfnamefont {Ernst~Jan}\ \bibnamefont
  {Vesseur}}, \ and\ \bibinfo {author} {\bibfnamefont {Albert}\ \bibnamefont
  {Polman}},\ }\bibfield  {title} {\enquote {\bibinfo {title}
  {Cathodoluminescence microscopy: optical imaging and spectroscopy with
  deep-subwavelength resolution},}\ }\href@noop {} {\bibfield  {journal}
  {\bibinfo  {journal} {MRS Bulletin}\ }\textbf {\bibinfo {volume} {40}},\
  \bibinfo {pages} {359--365} (\bibinfo {year} {2015})}\BibitemShut {NoStop}%
\bibitem [{\citenamefont {Kociak}\ \emph {et~al.}(2014)\citenamefont {Kociak},
  \citenamefont {St{\'e}phan}, \citenamefont {Gloter}, \citenamefont {Zagonel},
  \citenamefont {Tizei}, \citenamefont {Tenc{\'e}}, \citenamefont {March},
  \citenamefont {Blazit}, \citenamefont {Mahfoud}, \citenamefont {Losquin}
  \emph {et~al.}}]{kociakNanoOpticsEMReview}%
  \BibitemOpen
  \bibfield  {author} {\bibinfo {author} {\bibfnamefont {Mathieu}\ \bibnamefont
  {Kociak}}, \bibinfo {author} {\bibfnamefont {Odile}\ \bibnamefont
  {St{\'e}phan}}, \bibinfo {author} {\bibfnamefont {Alexandre}\ \bibnamefont
  {Gloter}}, \bibinfo {author} {\bibfnamefont {Luiz~F}\ \bibnamefont
  {Zagonel}}, \bibinfo {author} {\bibfnamefont {Luiz~HG}\ \bibnamefont
  {Tizei}}, \bibinfo {author} {\bibfnamefont {Marcel}\ \bibnamefont
  {Tenc{\'e}}}, \bibinfo {author} {\bibfnamefont {Katia}\ \bibnamefont
  {March}}, \bibinfo {author} {\bibfnamefont {Jean~Denis}\ \bibnamefont
  {Blazit}}, \bibinfo {author} {\bibfnamefont {Zackaria}\ \bibnamefont
  {Mahfoud}}, \bibinfo {author} {\bibfnamefont {Arthur}\ \bibnamefont
  {Losquin}},  \emph {et~al.},\ }\bibfield  {title} {\enquote {\bibinfo {title}
  {Seeing and measuring in colours: Electron microscopy and spectroscopies
  applied to nano-optics},}\ }\href@noop {} {\bibfield  {journal} {\bibinfo
  {journal} {Comptes Rendus Physique}\ }\textbf {\bibinfo {volume} {15}},\
  \bibinfo {pages} {158--175} (\bibinfo {year} {2014})}\BibitemShut {NoStop}%
\bibitem [{\citenamefont {Durham}\ \emph {et~al.}(2018)\citenamefont {Durham},
  \citenamefont {Ogletree},\ and\ \citenamefont
  {Barnard}}]{durhamScopeFoundry}%
  \BibitemOpen
  \bibfield  {author} {\bibinfo {author} {\bibfnamefont {Daniel~B}\
  \bibnamefont {Durham}}, \bibinfo {author} {\bibfnamefont {D~Frank}\
  \bibnamefont {Ogletree}}, \ and\ \bibinfo {author} {\bibfnamefont {Edward~S}\
  \bibnamefont {Barnard}},\ }\bibfield  {title} {\enquote {\bibinfo {title}
  {Scanning {Auger} spectromicroscopy using the {ScopeFoundry} software
  platform},}\ }\href@noop {} {\bibfield  {journal} {\bibinfo  {journal}
  {Surface and Interface Analysis}\ }\textbf {\bibinfo {volume} {50}},\
  \bibinfo {pages} {1174--1179} (\bibinfo {year} {2018})}\BibitemShut {NoStop}%
\bibitem [{\citenamefont {Barnard}()}]{ScopeFoundryWebsite}%
  \BibitemOpen
  \bibfield  {author} {\bibinfo {author} {\bibfnamefont {Edward~S}\
  \bibnamefont {Barnard}},\ }\href@noop {} {\enquote {\bibinfo {title}
  {{ScopeFoundry}: A python platform for controlling custom laboratory
  experiments and visualizing scientific data},}\ }\bibinfo {howpublished}
  {\url{http://www.scopefoundry.org/}}\BibitemShut {NoStop}%
\bibitem [{\citenamefont {Coenen}\ \emph {et~al.}(2011)\citenamefont {Coenen},
  \citenamefont {Vesseur},\ and\ \citenamefont
  {Polman}}]{coenenAPLAngleResolvedCL}%
  \BibitemOpen
  \bibfield  {author} {\bibinfo {author} {\bibfnamefont {Toon}\ \bibnamefont
  {Coenen}}, \bibinfo {author} {\bibfnamefont {Ernst Jan~R}\ \bibnamefont
  {Vesseur}}, \ and\ \bibinfo {author} {\bibfnamefont {Albert}\ \bibnamefont
  {Polman}},\ }\bibfield  {title} {\enquote {\bibinfo {title} {Angle-resolved
  cathodoluminescence spectroscopy},}\ }\href@noop {} {\bibfield  {journal}
  {\bibinfo  {journal} {Applied Physics Letters}\ }\textbf {\bibinfo {volume}
  {99}},\ \bibinfo {pages} {143103} (\bibinfo {year} {2011})}\BibitemShut
  {NoStop}%
\bibitem [{\citenamefont {Coenen}\ \emph {et~al.}(2014)\citenamefont {Coenen}
  \emph {et~al.}}]{coenenBookAngleResolvedCL}%
  \BibitemOpen
  \bibfield  {author} {\bibinfo {author} {\bibfnamefont {Toon}\ \bibnamefont
  {Coenen}} \emph {et~al.},\ }\href@noop {} {\emph {\bibinfo {title}
  {Angle-resolved cathodoluminescence nanoscopy}}}\ (\bibinfo  {publisher}
  {Universiteit van Amsterdam [Host]},\ \bibinfo {year} {2014})\BibitemShut
  {NoStop}%
\bibitem [{\citenamefont {Chaturvedi}\ \emph {et~al.}(2009)\citenamefont
  {Chaturvedi}, \citenamefont {Hsu}, \citenamefont {Kumar}, \citenamefont
  {Fung}, \citenamefont {Mabon},\ and\ \citenamefont
  {Fang}}]{chaturvedi2009CLDipoleFDTD}%
  \BibitemOpen
  \bibfield  {author} {\bibinfo {author} {\bibfnamefont {Pratik}\ \bibnamefont
  {Chaturvedi}}, \bibinfo {author} {\bibfnamefont {Keng~H}\ \bibnamefont
  {Hsu}}, \bibinfo {author} {\bibfnamefont {Anil}\ \bibnamefont {Kumar}},
  \bibinfo {author} {\bibfnamefont {Kin~Hung}\ \bibnamefont {Fung}}, \bibinfo
  {author} {\bibfnamefont {James~C}\ \bibnamefont {Mabon}}, \ and\ \bibinfo
  {author} {\bibfnamefont {Nicholas~X}\ \bibnamefont {Fang}},\ }\bibfield
  {title} {\enquote {\bibinfo {title} {Imaging of plasmonic modes of silver
  nanoparticles using high-resolution cathodoluminescence spectroscopy},}\
  }\href@noop {} {\bibfield  {journal} {\bibinfo  {journal} {ACS Nano}\
  }\textbf {\bibinfo {volume} {3}},\ \bibinfo {pages} {2965--2974} (\bibinfo
  {year} {2009})}\BibitemShut {NoStop}%
\bibitem [{\citenamefont {Osorio}\ \emph {et~al.}(2015)\citenamefont {Osorio},
  \citenamefont {Coenen}, \citenamefont {Brenny}, \citenamefont {Polman},\ and\
  \citenamefont {Koenderink}}]{polmanBullseyeCLPolarimetry}%
  \BibitemOpen
  \bibfield  {author} {\bibinfo {author} {\bibfnamefont {Clara~I}\ \bibnamefont
  {Osorio}}, \bibinfo {author} {\bibfnamefont {Toon}\ \bibnamefont {Coenen}},
  \bibinfo {author} {\bibfnamefont {Benjamin~JM}\ \bibnamefont {Brenny}},
  \bibinfo {author} {\bibfnamefont {Albert}\ \bibnamefont {Polman}}, \ and\
  \bibinfo {author} {\bibfnamefont {A~Femius}\ \bibnamefont {Koenderink}},\
  }\bibfield  {title} {\enquote {\bibinfo {title} {Angle-resolved
  cathodoluminescence imaging polarimetry},}\ }\href@noop {} {\bibfield
  {journal} {\bibinfo  {journal} {ACS Photonics}\ }\textbf {\bibinfo {volume}
  {3}},\ \bibinfo {pages} {147--154} (\bibinfo {year} {2015})}\BibitemShut
  {NoStop}%
\bibitem [{\citenamefont {Kr{\"u}ger}\ \emph {et~al.}(2012)\citenamefont
  {Kr{\"u}ger}, \citenamefont {Schenk}, \citenamefont {F{\"o}rster},\ and\
  \citenamefont {Hommelhoff}}]{hommelhoff2012opticalFieldEmission}%
  \BibitemOpen
  \bibfield  {author} {\bibinfo {author} {\bibfnamefont {Michael}\ \bibnamefont
  {Kr{\"u}ger}}, \bibinfo {author} {\bibfnamefont {Markus}\ \bibnamefont
  {Schenk}}, \bibinfo {author} {\bibfnamefont {Michael}\ \bibnamefont
  {F{\"o}rster}}, \ and\ \bibinfo {author} {\bibfnamefont {Peter}\ \bibnamefont
  {Hommelhoff}},\ }\bibfield  {title} {\enquote {\bibinfo {title} {Attosecond
  physics in photoemission from a metal nanotip},}\ }\href@noop {} {\bibfield
  {journal} {\bibinfo  {journal} {Journal of Physics B: Atomic, Molecular and
  Optical Physics}\ }\textbf {\bibinfo {volume} {45}},\ \bibinfo {pages}
  {074006} (\bibinfo {year} {2012})}\BibitemShut {NoStop}%
\bibitem [{\citenamefont {Piglosiewicz}\ \emph {et~al.}(2014)\citenamefont
  {Piglosiewicz}, \citenamefont {Schmidt}, \citenamefont {Park}, \citenamefont
  {Vogelsang}, \citenamefont {Gro{\ss}}, \citenamefont {Manzoni}, \citenamefont
  {Farinello}, \citenamefont {Cerullo},\ and\ \citenamefont
  {Lienau}}]{lienau2014cepPhotoemission}%
  \BibitemOpen
  \bibfield  {author} {\bibinfo {author} {\bibfnamefont {Bj{\"o}rn}\
  \bibnamefont {Piglosiewicz}}, \bibinfo {author} {\bibfnamefont {Slawa}\
  \bibnamefont {Schmidt}}, \bibinfo {author} {\bibfnamefont {Doo~Jae}\
  \bibnamefont {Park}}, \bibinfo {author} {\bibfnamefont {Jan}\ \bibnamefont
  {Vogelsang}}, \bibinfo {author} {\bibfnamefont {Petra}\ \bibnamefont
  {Gro{\ss}}}, \bibinfo {author} {\bibfnamefont {Cristian}\ \bibnamefont
  {Manzoni}}, \bibinfo {author} {\bibfnamefont {Paolo}\ \bibnamefont
  {Farinello}}, \bibinfo {author} {\bibfnamefont {Giulio}\ \bibnamefont
  {Cerullo}}, \ and\ \bibinfo {author} {\bibfnamefont {Christoph}\ \bibnamefont
  {Lienau}},\ }\bibfield  {title} {\enquote {\bibinfo {title} {Carrier-envelope
  phase effects on the strong-field photoemission of electrons from metallic
  nanostructures},}\ }\href@noop {} {\bibfield  {journal} {\bibinfo  {journal}
  {Nature Photonics}\ }\textbf {\bibinfo {volume} {8}},\ \bibinfo {pages} {37}
  (\bibinfo {year} {2014})}\BibitemShut {NoStop}%
\bibitem [{\citenamefont {England}\ \emph {et~al.}(2014)\citenamefont
  {England}, \citenamefont {Noble}, \citenamefont {Bane}, \citenamefont
  {Dowell}, \citenamefont {Ng}, \citenamefont {Spencer}, \citenamefont
  {Tantawi}, \citenamefont {Wu}, \citenamefont {Byer}, \citenamefont {Peralta},
  \citenamefont {Soong}, \citenamefont {Chang}, \citenamefont {Montazeri},
  \citenamefont {Wolf}, \citenamefont {Cowan}, \citenamefont {Dawson},
  \citenamefont {Gai}, \citenamefont {Hommelhoff}, \citenamefont {Huang},
  \citenamefont {Jing}, \citenamefont {McGuinness}, \citenamefont {Palmer},
  \citenamefont {Naranjo}, \citenamefont {Rosenzweig}, \citenamefont {Travish},
  \citenamefont {Mizrahi}, \citenamefont {Schachter}, \citenamefont {Sears},
  \citenamefont {Werner},\ and\ \citenamefont {Yoder}}]{DLA}%
  \BibitemOpen
  \bibfield  {author} {\bibinfo {author} {\bibfnamefont {R.~Joel}\ \bibnamefont
  {England}}, \bibinfo {author} {\bibfnamefont {Robert~J.}\ \bibnamefont
  {Noble}}, \bibinfo {author} {\bibfnamefont {Karl}\ \bibnamefont {Bane}},
  \bibinfo {author} {\bibfnamefont {David~H.}\ \bibnamefont {Dowell}}, \bibinfo
  {author} {\bibfnamefont {Cho-Kuen}\ \bibnamefont {Ng}}, \bibinfo {author}
  {\bibfnamefont {James~E.}\ \bibnamefont {Spencer}}, \bibinfo {author}
  {\bibfnamefont {Sami}\ \bibnamefont {Tantawi}}, \bibinfo {author}
  {\bibfnamefont {Ziran}\ \bibnamefont {Wu}}, \bibinfo {author} {\bibfnamefont
  {Robert~L.}\ \bibnamefont {Byer}}, \bibinfo {author} {\bibfnamefont {Edgar}\
  \bibnamefont {Peralta}}, \bibinfo {author} {\bibfnamefont {Ken}\ \bibnamefont
  {Soong}}, \bibinfo {author} {\bibfnamefont {Chia-Ming}\ \bibnamefont
  {Chang}}, \bibinfo {author} {\bibfnamefont {Behnam}\ \bibnamefont
  {Montazeri}}, \bibinfo {author} {\bibfnamefont {Stephen~J.}\ \bibnamefont
  {Wolf}}, \bibinfo {author} {\bibfnamefont {Benjamin}\ \bibnamefont {Cowan}},
  \bibinfo {author} {\bibfnamefont {Jay}\ \bibnamefont {Dawson}}, \bibinfo
  {author} {\bibfnamefont {Wei}\ \bibnamefont {Gai}}, \bibinfo {author}
  {\bibfnamefont {Peter}\ \bibnamefont {Hommelhoff}}, \bibinfo {author}
  {\bibfnamefont {Yen-Chieh}\ \bibnamefont {Huang}}, \bibinfo {author}
  {\bibfnamefont {Chunguang}\ \bibnamefont {Jing}}, \bibinfo {author}
  {\bibfnamefont {Christopher}\ \bibnamefont {McGuinness}}, \bibinfo {author}
  {\bibfnamefont {Robert~B.}\ \bibnamefont {Palmer}}, \bibinfo {author}
  {\bibfnamefont {Brian}\ \bibnamefont {Naranjo}}, \bibinfo {author}
  {\bibfnamefont {James}\ \bibnamefont {Rosenzweig}}, \bibinfo {author}
  {\bibfnamefont {Gil}\ \bibnamefont {Travish}}, \bibinfo {author}
  {\bibfnamefont {Amit}\ \bibnamefont {Mizrahi}}, \bibinfo {author}
  {\bibfnamefont {Levi}\ \bibnamefont {Schachter}}, \bibinfo {author}
  {\bibfnamefont {Christopher}\ \bibnamefont {Sears}}, \bibinfo {author}
  {\bibfnamefont {Gregory~R.}\ \bibnamefont {Werner}}, \ and\ \bibinfo {author}
  {\bibfnamefont {Rodney~B.}\ \bibnamefont {Yoder}},\ }\bibfield  {title}
  {\enquote {\bibinfo {title} {Dielectric laser accelerators},}\ }\href
  {\doibase 10.1103/RevModPhys.86.1337} {\bibfield  {journal} {\bibinfo
  {journal} {Reviews of Modern Physics}\ }\textbf {\bibinfo {volume} {86}},\
  \bibinfo {pages} {1337--1389} (\bibinfo {year} {2014})}\BibitemShut {NoStop}%
\end{thebibliography}%

\end{document}